\documentclass{aa}
\usepackage{graphics}
\usepackage{times}

\newcommand{\mv}{\mbox{$M_{V}$}}

\newcommand{\logt}{\mbox{$\log(t/{\rm yr})$}}
\newcommand{\Msun}{\mbox{$M_{\odot}$}}

\newcommand{\sub}[1]{\mbox{$_{\rm #1}$}}

\newcommand{\Mhef}{\mbox{$M\sub{Hef}$}}
\newcommand{\Mup}{\mbox{$M\sub{up}$}}

\newcommand{\Teff}{\mbox{$T\sub{eff}$}}
\newcommand{\logTe}{\mbox{$\log T\sub{eff}$}}
\newcommand{\logL}{\mbox{$\log(L/L_{\odot})$}}
\newcommand{\diff}{\mbox{d}}

\newcommand{\reffig}[1]{Fig.\ \protect\ref{#1}}
\newcommand{\reftab}[1]{Table \protect\ref{#1}}
\newcommand{\refsec}[1]{Section \protect\ref{#1}}

\begin{document}

	\thesaurus{08(08.05.3, 08.09.3, 08.08.1, 08.12.2) }

        \title{Evolutionary tracks and isochrones for 
low- and intermediate-mass stars: from 0.15 to 7 $M_{\odot}$, 
and from $Z=0.0004$ to $0.03$.}

	\author{L\'eo Girardi$^{1,2,3}$, Alessandro Bressan$^4$,
Gianpaolo Bertelli$^{1,5}$, and Cesare Chiosi$^1$}
	\institute{
$^1$ Dipartimento di Astronomia, Universit\`a di Padova, 
	Vicolo dell'Osservatorio 5, I-35122 Padova, Italy \\
$^2$ Max-Planck-Institut f\"ur Astrophysik, 
	Karl-Schwarzschild-Str.~1, D-85740 Garching bei M\"unchen,
	Germany\\
$^3$ Instituto de F\'\i sica, Universidade Federal do Rio Grande do Sul, 
	Av.\ Bento Gon\c calves 9500, 
	91501-970 Porto Alegre RS, Brazil \\
$^4$ Osservatorio Astronomico di Padova, Vicolo dell'Osservatorio 5,
	I-35122 Padova, Italy \\
$^5$ Consiglio Nazionale delle Ricerche (CNR), Italy
	}

	\titlerunning{Evolutionary tracks and isochrones}
	\authorrunning{Girardi et al.}

	\offprints{L\'eo Girardi \\ 
e-mail: Lgirardi@pd.astro.it }

	\date{Submitted to Astronomy \& Astrophysics Supplement Series}

	\maketitle

	\begin{abstract}
We present a large grid of stellar evolutionary tracks, which are
suitable to modelling star clusters and galaxies by means of
population synthesis.  The tracks are presented for the initial
chemical compositions $[Z=0.0004, Y=0.23]$, $[Z=0.001, Y=0.23]$, 
$[Z=0.004, Y=0.24]$,
$[Z=0.008, Y=0.25]$, $[Z=0.019, Y=0.273]$ (solar composition), and
$[Z=0.03, Y=0.30]$.  They are computed with updated opacities and
equation of state, and a moderate amount of convective overshoot.  The
range of initial masses goes from $0.15~M_{\odot}$ to $7~M_{\odot}$,
and the evolutionary phases extend from the zero age main sequence
(ZAMS) till either the thermally pulsing AGB regime or carbon
ignition. We also present an additional set of models with solar 
composition, computed using the classical Schwarzschild's criterion
for convective boundaries. From all these tracks, we derive the
theoretical isochrones in the Johnson-Cousins $UBVRIJHK$ broad-band
photometric system.

\keywords{stars: evolution -- stars: interiors -- 
Hertz\-sprung--Russel (HR) diagram -- stars: low mass}

\end{abstract}  

\section{ Introduction }

Models of evolutionary population synthesis require the computation of
large data-bases of stellar evolutionary tracks, being as far as
possible extended in their interval of initial masses and
metallicities, covering the main evolutionary phases, and adopting a 
homogeneous and updated input physics.

One of the most popular of these sets is that from the Padova
group (Bressan et al.\ 1993; Fagotto et al.\
1994ab; Bertelli et al.\ 1994, and references therein). The complete
data-base covers a very large range of stellar masses (typically from
$0.6$ to $120$~\Msun) and metallicities (from $Z=0.0004$ to
0.05). The input physics is homogeneous for all stellar 
tracks in the grids. Main characteristics of these models are the
adoption of OPAL opacities, a constant helium-to-metal enrichment
ratio $\Delta Y/\Delta Z$, moderate convective overshooting from
convective cores, and mass loss from massive stars. This data-base was
extended to very low metallicities (i.e.\ $Z=0.0001$) by Girardi
et al.\ (1996a).

Since 1994, many other evolutionary tracks have been computed by us,
following some major updatings in the input physics. Some of the new
tracks were intended only for testing the effects of these
modifications, such as e.g.\ those discussed in Girardi et al.\
(1996b). The main novelties with respect to the Bertelli et al.\ (1994)
tracks are, essentially, the adoption of an improved equation of state
(Straniero 1988; Girardi et al.\ 1996a; Mihalas et al.\ 1990 and
references therein), and the new low-temperature opacities from
Alexander \& Ferguson (1994). Eventually, complete sets of
evolutionary tracks have been computed, thus generating a
data-base which is comparable to the previous one in terms of 
the large coverage of mass and metallicity ranges.

This paper aims to present this new data-base of stellar evolutionary
tracks and isochrones, and make it available to the users. The
tracks presented here are limited to the interval of low- and
intermediate mass stars (i.e.\ from 0.15 to 7~\Msun), and to 6
values of metallicity, from $Z=0.0004$ to $Z=0.03$. Further extensions
of this data-base will be provided in subsequent papers.

The new evolutionary tracks have already been used in several studies in
the last years. They are the starting point for the detailed thermally
pulsing AGB tracks of Marigo (1998ab) and Marigo et al.\ (1998,
1999a). The stellar chemical yields have been computed by Marigo
(1998a) and Marigo et al.\ (1999b). An extended sample of the
low-mass ZAMS models has been used by H\o g et al.\ (1998) and Pagel
\& Portinari (1998) in order to access the helium-to-metal enrichment
ratio, $\Delta Y/\Delta Z$, for the local stars sampled by the
astrometric satellite HIPPARCOS. Girardi (1996) and Girardi \&
Bertelli (1998) discussed the behaviour of the $B-V$ and $V-K$
integrated colours of single-burst stellar populations derived from
the present tracks.  Pasquini et al.\ (1999) used the evolution of 
the momentum of inertia of these models to interpretate observations
of Disc stars. Girardi et al.\ (1998) and Girardi (1999)
generated synthetic colour-magnitude diagrams (CMD), finding
substructures on the clump of red giants of different galaxy
models. Finally, the present isochrones were used by Carraro
et al.\ (1999ab) to derive the ages of the oldest known open
clusters.

The plane of this paper is as follows: \refsec{sec_input} presents the
input physics of the models; \refsec{sec_tracks} introduces the
stellar tracks and discuss their main characteristics;
\refsec{sec_rgbagb} describes the mass-loss on the RGB and synthetic
TP-AGB evolution; 
\refsec{sec_isochrones} describes the derived isochrones. Finally
\refsec{sec_remarks} comments on the compatibility of present models
with the Bertelli et al.\ (1994) data-base.

\section{Input physics}
\label{sec_input}

\subsection{Initial chemical composition}
\label{sec_chemic}

Stellar models are assumed to be chemically homogeneous when they
settle on the zero age main sequence (ZAMS). The helium and metal mass
fractions, $Y$ and $Z$, are chosen according to some fixed $Y(Z)$
relation. In the present models, the values of $[Z=0.0004, Y=0.23]$, 
$[Z=0.001, Y=0.23]$,
$[Z=0.004, Y=0.24]$, and $[Z=0.008, Y=0.25]$, were chosen in order to
coincide with the choices previously adopted in Bertelli et al.\
(1994, and references therein).  On the other hand, the set with solar 
composition was computed with $[Z=0.019, Y=0.273]$, since this
value of $Y$ was fixed by the calibration of the solar model (see
\refsec{sec_sun} below).  From these 5 pairs of $[Z, Y]$ values, we
obtain an helium-to-metal enrichment relation which is 
$Y\simeq0.23+2.25\,Z$. We then decide to adopt this mean relation for
super-solar metallicities. It gives origin to the values of $[Z=0.03,
Y=0.30]$ adopted in our set of highest metallicity.

For each value of $Z$, the fractions of different metals follow a
solar-scaled distribution, as compiled by 
Grevesse \& Noels (1993) and adopted
in the OPAL opacity tables. The ratio between abundances of different
isotopes is according to Anders \& Grevesse (1989).

\subsection{Opacities}
\label{sec_opac}

The radiative opacities are from the OPAL group
(Rogers \& Iglesias 1992; Iglesias \& Rogers 1993) for temperatures
higher than $\log(T/{\rm K})=4.1$, and from Alexander \& Ferguson
(1994) for $\log(T/{\rm K})<4.0$. In the temperature interval 
$4.0<\log(T/{\rm K})<4.1$, a
linear interpolation between the opacities derived from both
sources is adopted. We remind the reader that both opacities sources 
provide values in good agreement in this temperature interval; the 
relative differences in opacities are typically lower than 5 percent 
(see Alexander \& Ferguson 1994).

The conductive opacities of electron-degenerate matter are
from Hubbard \& Lampe (1969). We compared the tracks obtained with 
this prescription with more recent ones (Salasnich et al., 
1999) which use the Itoh et al.\ (1983) formulas.
No significant differences in the evolutionary features turned out.

\subsection{Equation of state}
\label{sec_eos}

The equation of state (EOS) for temperatures higher than $10^7$~K is
that of a fully-ionized gas, including electron degeneracy in the way
described by Kippenhahn et al.\ (1965). The effect of Coulomb
interactions between the gas particles at high densities is
introduced following the prescription by Straniero (1988). The latter
was however adapted to the general case of a multiple-component
plasma, as described in the appendix of Girardi et al.\ (1996a).

For temperatures lower than $10^7$~K, we adopt the detailed ``MHD''
EOS of Mihalas et al.\ (1990, and references therein). The free-energy
minimization technique used to derive thermodynamical quantities and
derivatives for any input chemical composition, is described in detail
by Hummer \& Mihalas (1988), D\"appen et al.\ (1988), and Mihalas et
al.\ (1988). In our cases, we explicitly calculated EOS tables for all
the $Z$ values of our tracks, using the Mihalas et al.\ (1990)
code. To save computer time, we consider only the four most abundant
metal species, i.e.\ C, N, O, and Ne. For each $Z$, EOS tables for
several closely spaced values of $Y$ were computed, in order to cover the
range of surface helium composition found in the evolutionary models
before and after the first and second dredge-up events.

Alternatively, we computed some tracks using a much simpler EOS, where
the ionization equilibrium and thermo-dynamical quantities were derived
by solving a simple set of Saha equations for a H+He mixture (Baker \&
Kippenhahn 1962). Comparison of these tracks with those obtained with
the MHD EOS revealed that no significant differences in effective
temperature or luminosity arise for dwarf stars of mass higher than
about 0.7~\Msun, or for giant stars of any mass. This is so because
only dwarf stars of lower mass present the dense and cold envelopes in
which the non-ideal effects included in the MHD EOS become
important. Moreover, only in the lowest-mass stars the surface
temperatures are low enough so that the H$_2$ molecule is formed,
which has dramatic consequences for their internal structures (see
e.g.\ Copeland et al.\ 1970). Therefore, the use of the MHD EOS is
essential for our aim of computing stellar models with masses lower
than 0.6~\Msun\ (see also Girardi et al.\ 1996b).

\subsection{Reaction rates and neutrino losses}
\label{sec_rates}

The network of nuclear reactions we use involves all the important
reactions of the pp and CNO chains, and the most important
alpha-capture reactions for elements as heavy as Mg (see Bressan et
al.\ 1993 and Maeder 1983 for details). 

The reaction rates are from the compilation of Caughlan \& Fowler
(1988), but for $^{17}{\rm O}({\rm
p},\alpha)^{14}{\rm N}$ and $^{17}{\rm O}({\rm p},\gamma)^{18}{\rm
F}$, for which we use the more recent determinations by 
Landr\'e et al.\ (1990). The
uncertain $^{12}$C($\alpha,\gamma$)$^{16}$O rate was set to 1.7 times
the values given by Caughlan \& Fowler (1988), as indicated by the
study of Weaver \& Woosley (1993) on the nucleosynthesis by massive
stars.  The electron screening factors for all reactions are those
from Graboske et al.\ (1973).

The energy losses by pair, plasma, and bremsstrahlung neutrinos,
important in the electron degenerate stellar cores, are taken from
Munakata et al.\ (1985) and Itoh \& Kohyama (1983).

\subsection{Convection}
\label{sec_conv}

The energy transport in the outer convection zone is described
according to the mixing-length theory of B\"ohm-Vitense (1958). The
mixing length parameter $\alpha$ is calibrated 
by means of the solar model (see \refsec{sec_sun} below).

The extension of convective boundaries is estimated by means of an
algorithm which takes into account overshooting from the borders of
both core and envelope convective zones. The formalism is fully
described in Bressan et al.\ (1981) and Alongi et al.\ (1991).
The main parameter describing
overshooting is its extent $\Lambda_{\rm c}$ {\em across}
the border of the convective zone, expressed in units of pressure
scale heigth. Importantly, this parameter in the Bressan et al.\ (1981)
formalism is not equivalent to others found in literature. For
instance, the overshooting scale defined by $\Lambda_{\rm c}=0.5$ in
the Padova formalism roughly corresponds to the 0.25 pressure scale
heigth {\em above} the convective border, adopted by the Geneva group
(Meynet et al.\ 1994 and references therein) 
to describe the same physical phenomenum, i.e.\
$\Lambda^{\rm G}_{\rm c}=0.25$. The non-equivalency of the parameters
used to describe convective overshooting by different groups, has been
a recurrent source of misunderstanding in the literature.

We adopt the following prescription for the
parameter $\Lambda_{\rm c}$ as a function of stellar mass:
	\begin{itemize}
	\item
$\Lambda_{\rm c}$ is set to zero for stellar masses $M\le1.0$~\Msun,
in order to avoid the development of a small central convective zone
in the solar model, which would persist up to the present age of 4.6
Gyr. 
	\item 
For $M\ge1.5$~\Msun, we adopt $\Lambda_{\rm c}=0.5$,
i.e.\ a moderate amount of overshoting, which coincides with the
values adopted in the previous Bertelli et al.\ (1994) models. 
	\item
In the
range $1.0<M<1.5$~\Msun, we adopt a gradual increase of the overshoot
efficiency with mass, i.e.\ $\Lambda_{\rm c}=M/\Msun - 1.0$. This
because the calibration of the overshooting efficiency in this mass 
range is still very uncertain due to the scarcity of stellar data 
provided by the oldest intermediate-age and old open clusters. 
Some works (eg.\ Aparicio et al.\ 1990), however, indicate that this 
efficiency should be lower than in intermediate-mass stars.
	\end{itemize} 

In the stages of core helium burning (CHeB), the value $\Lambda_{\rm
c}=0.5$ is used for all stellar masses. This amount of overshooting
dramatically reduces the extent of the breathing pulses of convection
found in the late phases of CHeB (see Chiosi et al.\ 1992).

Overshooting at the lower boundary of
convective envelopes is also considered. 
The calibration of the solar model required 
an envelope overshooting not higher than 0.25 pressure scale
heigths. This value of $\Lambda_{\rm e}=0.25$ (see Alongi et al.\
1991, for a description of the formalism) was then adopted for the
stars with $0.6\le(M/\Msun)<2.0$, whereas $\Lambda_{\rm e}=0$ was
adopted for $M<0.6$~\Msun. On the other hand, low values of
$\Lambda_{\rm e}$ lead to the almost complete suppression of the
Cepheid loops in intermediate-mass models (Alongi et al.\ 1991). 
Therefore, for
$M>2.5$~\Msun\ a value of $\Lambda_{\rm e}=0.7$ was assumed as in
Bertelli et al.\ (1994). Finally, for masses between 2.0 and
$2.5$~\Msun, $\Lambda_{\rm e}$ was let to increase gradually from 0.25
to 0.7.

We are well aware that the present prescription for the overshooting
parameters seems not to fit on the ideals of simplicity and
homogeneity one would like to find in such large sets of evolutionary
tracks. However, they represent a pragmatic choice, since the
prescriptions previously adopted were not satisfactory in many
details.

	\begin{figure*}
	\resizebox{12cm}{!}{\includegraphics{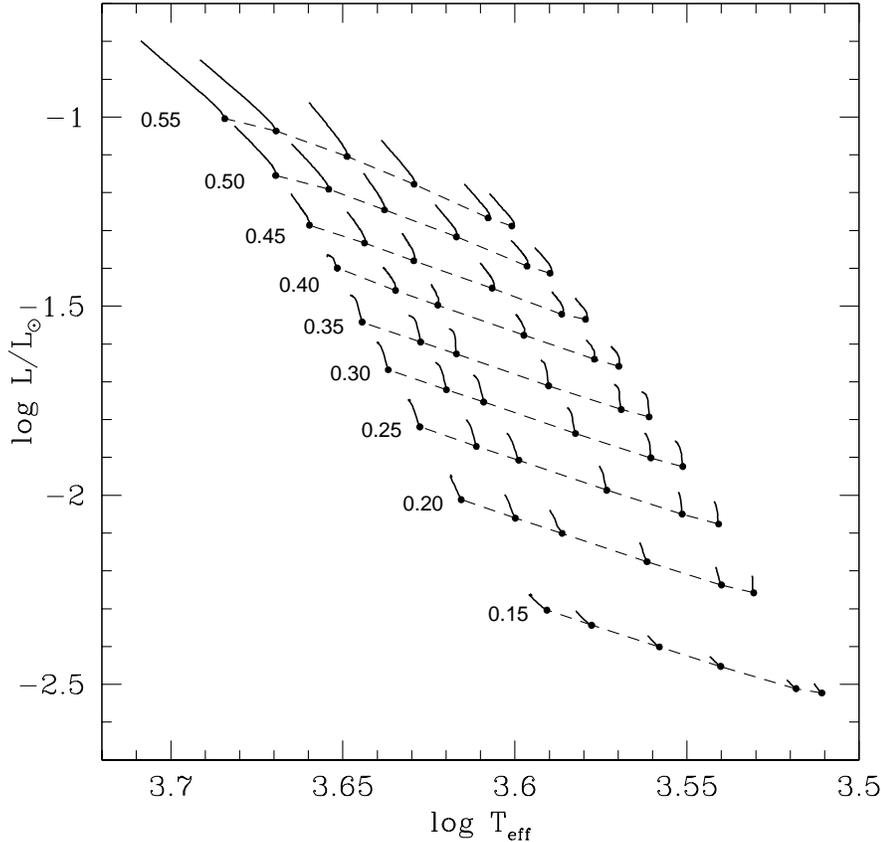}}
	\hfill
	\parbox[b]{55mm}{
        \caption{
Evolutionary tracks in the HR diagram, for the models with mass lower
than 0.6~\Msun, from the ZAMS up to an age of 25 Gyr. For each star, 
the evolution starts at the full dot and procedes at increasing 
luminosity, along the continuous lines. For each stellar mass 
indicated at the left part of the plot, six tracks are presented,
for the metallicity values $Z=0.0004$, 0.001, 0.004, 0.008, 0.019, 
0.030 (along the dashed line, from left to right).}
	\label{fig_bassa}
	}
	\end{figure*} 

\subsection{Calibration of the solar model}
\label{sec_sun}

The calibration of the solar model is an essential step in the
computation of our evolutionary tracks. Some of the parameters found
in the solar model are subsequentely adopted in all the stellar models
of our data-base.

We adopt for the Sun the metallicity of $Z=0.019$, i.e.\ a value
almost identical to the $Z_\odot=0.01886$ favoured by Anders \&
Grevesse (1989). Several 1~\Msun\ models, for different values of
mixing-length parameter $\alpha$ and helium content $Y_\odot$, are let
to evolve up to the age of 4.6~Gyr. From this series of models, we are
able to single out the pair of $[\alpha, Y_\odot]$ which allows
for a simultaneous match of the present-day solar radius and
luminosity, $R_\odot$ and $L_\odot$.

An additional constraint 
for the solar model comes from the heliosismological
determination of the depth of the solar convection zone, of 
$0.287\pm0.003\;R_{\odot}$ (Christensen-Dalsgaard et al.\ 1991). 
It corresponds to a radius of $R_{\rm c}=0.713\;R_{\odot}$ for
the lower boundary of the convective envelope. The
adoption of a overshooting parameter of $\Lambda_{\rm e}=0.7$, like in
Bertelli et al.\ (1994), would lead to a solar model with too a deep
convection zone, namely with $R_{\rm c}=0.680\;R_{\odot}$. 
This parameter was then reduced to $\Lambda_{\rm
e}=0.25$, which allows a reasonable reproduction of the observed value
of $R_{\rm c}$.

Our final solar model reproduces well the Sun $R_\odot$, $L_\odot$,
and $R_{\rm c}$ values. From
this model, we derive the values of $\alpha=1.68$, $Y_\odot=0.273$,
and $\Lambda_{\rm e}=0.25$, used in other stellar models as described
in the previous subsections.

\begin{figure*}
\begin{minipage}{0.45\textwidth} \noindent a)
\resizebox{\hsize}{!}{\includegraphics{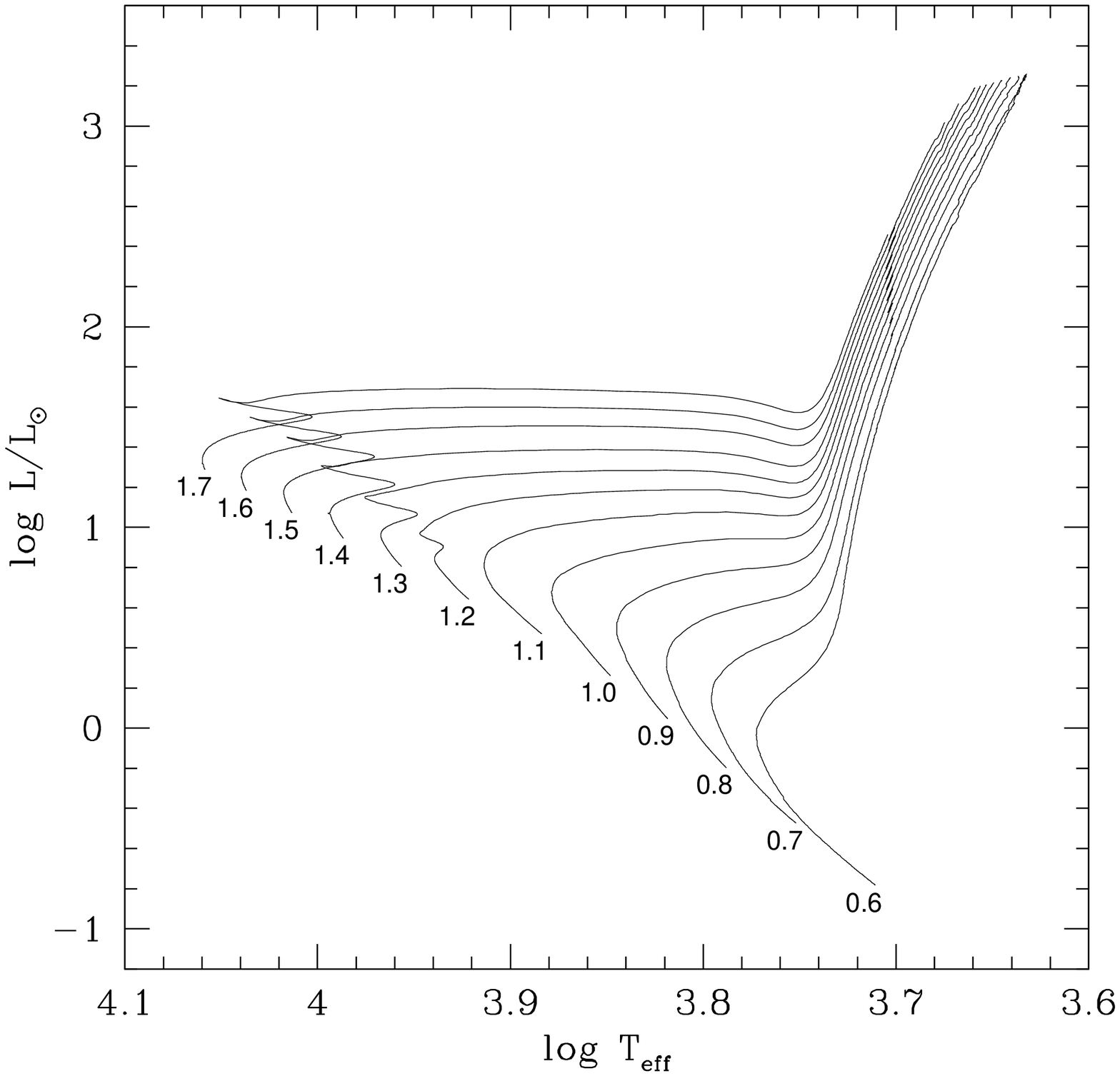}}
\end{minipage} 
\hfill
\begin{minipage}{0.45\textwidth} \noindent b) 
\resizebox{\hsize}{!}{\includegraphics{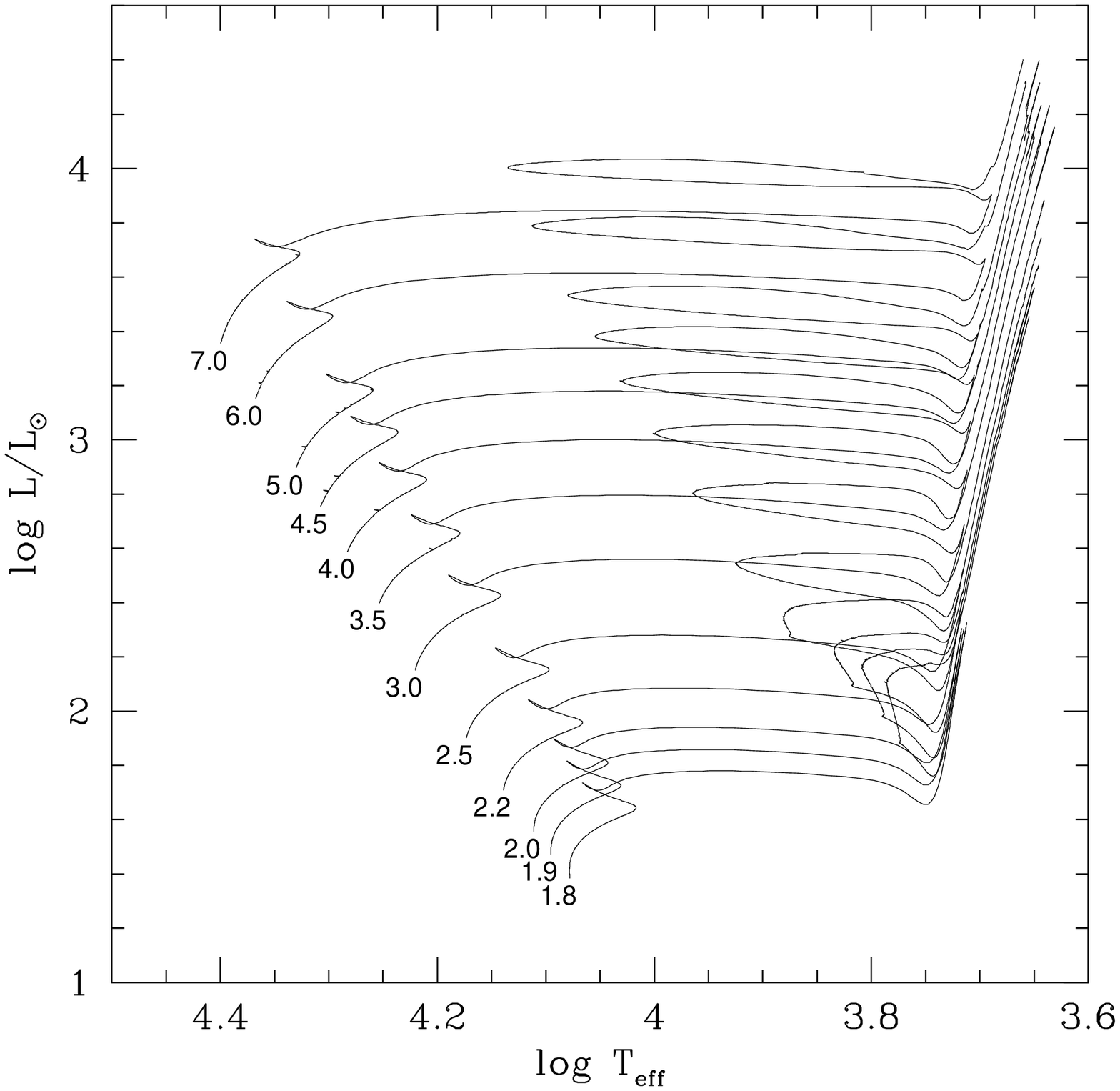}}
\end{minipage} 
\begin{minipage}{0.45\textwidth} \noindent c)
\resizebox{\hsize}{!}{\includegraphics{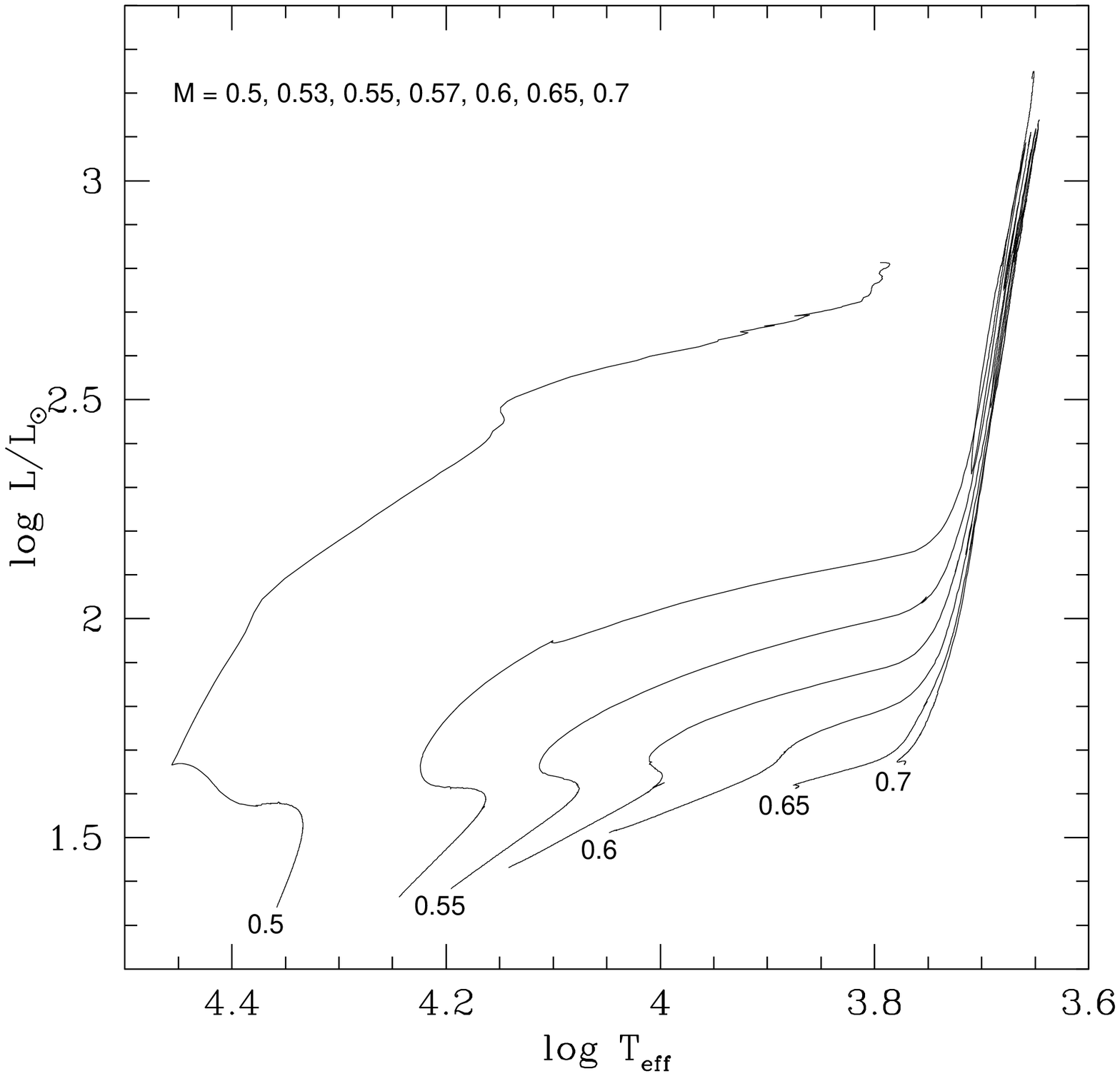}}
\end{minipage} 
\hfill
\begin{minipage}{0.45\textwidth} \noindent d)
\resizebox{\hsize}{!}{\includegraphics{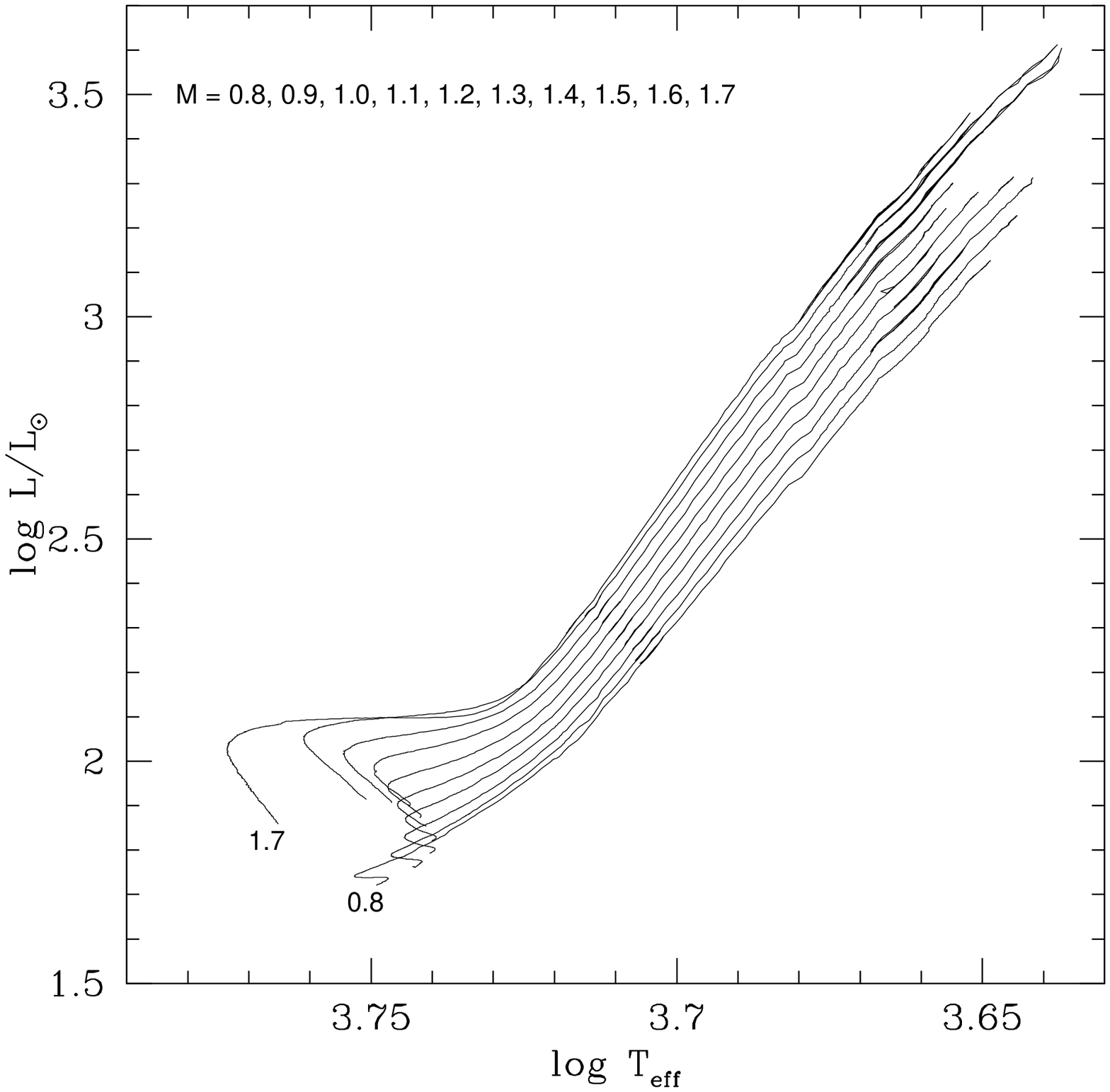}}
\end{minipage} 
\caption{  
Evolutionary tracks in the HR diagram, for the composition $[Z=0.0004,
Y=0.23]$. For most tracks of low-mass stars up to the RGB-tip 
(panel a), and intermediate-mass ones up to the TP-AGB (panel b), 
the stellar mass (in \Msun) is indicated 
at the initial point of the evolution. 
For the low-mass tracks from the ZAHB up to the TP-AGB (panels 
c and d), we indicate the complete range of stellar masses in the 
upper part of the plots.}
\label{hrd_z0004}
\end{figure*} 

\begin{figure*}
\begin{minipage}{0.45\textwidth} \noindent a)
\resizebox{\hsize}{!}{\includegraphics{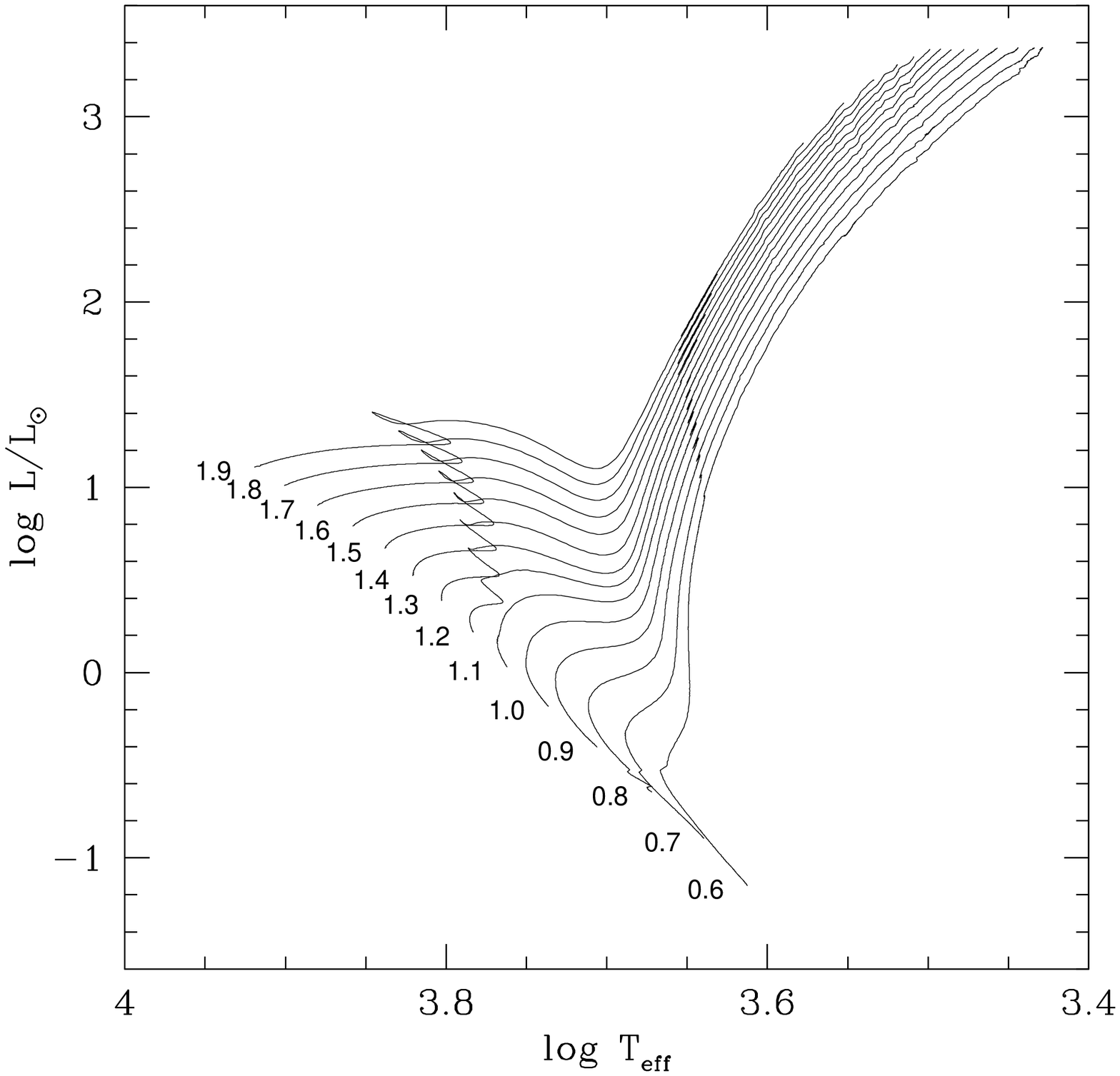}}
\end{minipage} 
\hfill
\begin{minipage}{0.45\textwidth} \noindent b) 
\resizebox{\hsize}{!}{\includegraphics{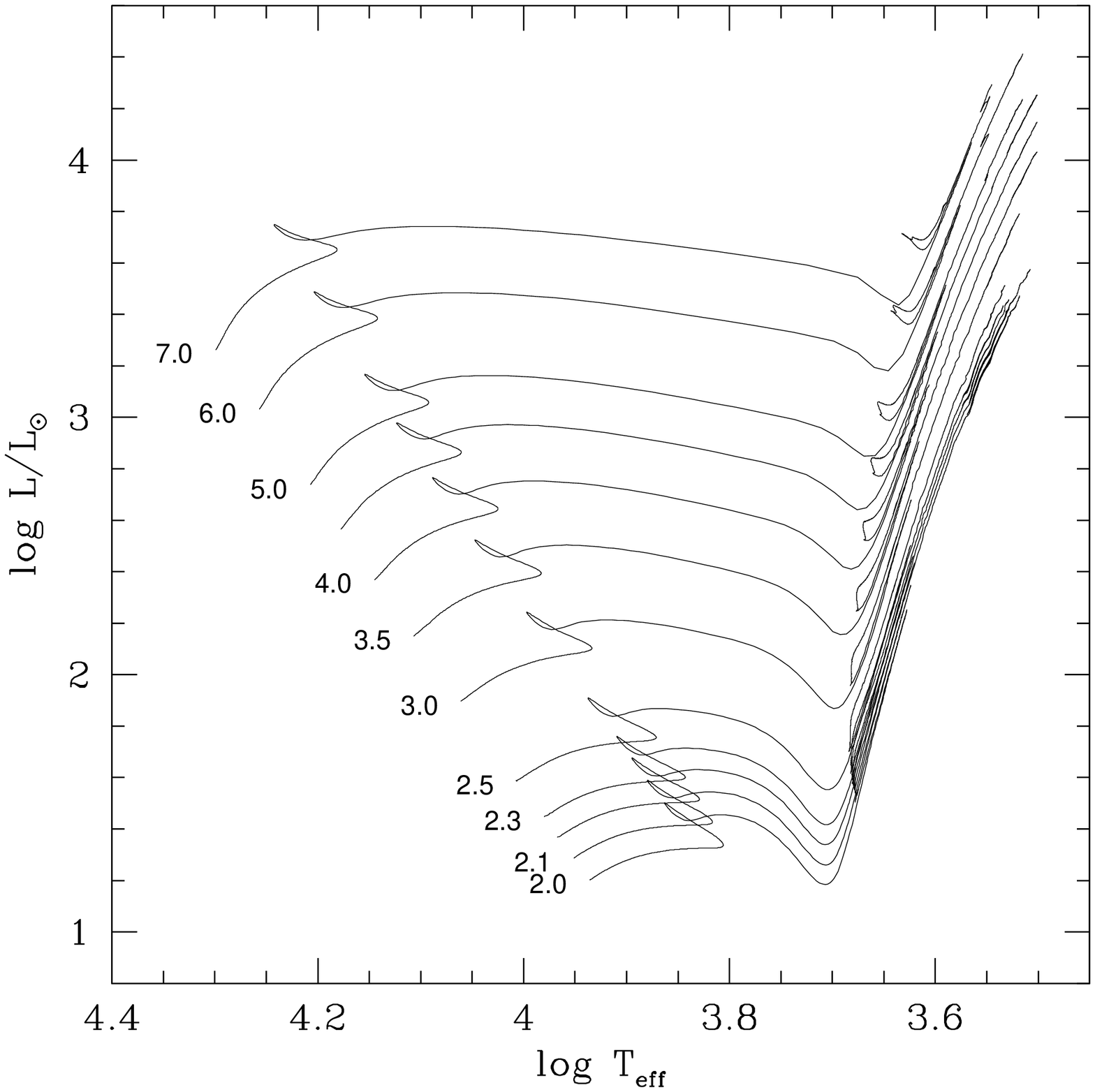}}
\end{minipage} 
\begin{minipage}{0.45\textwidth} \noindent c)
\resizebox{\hsize}{!}{\includegraphics{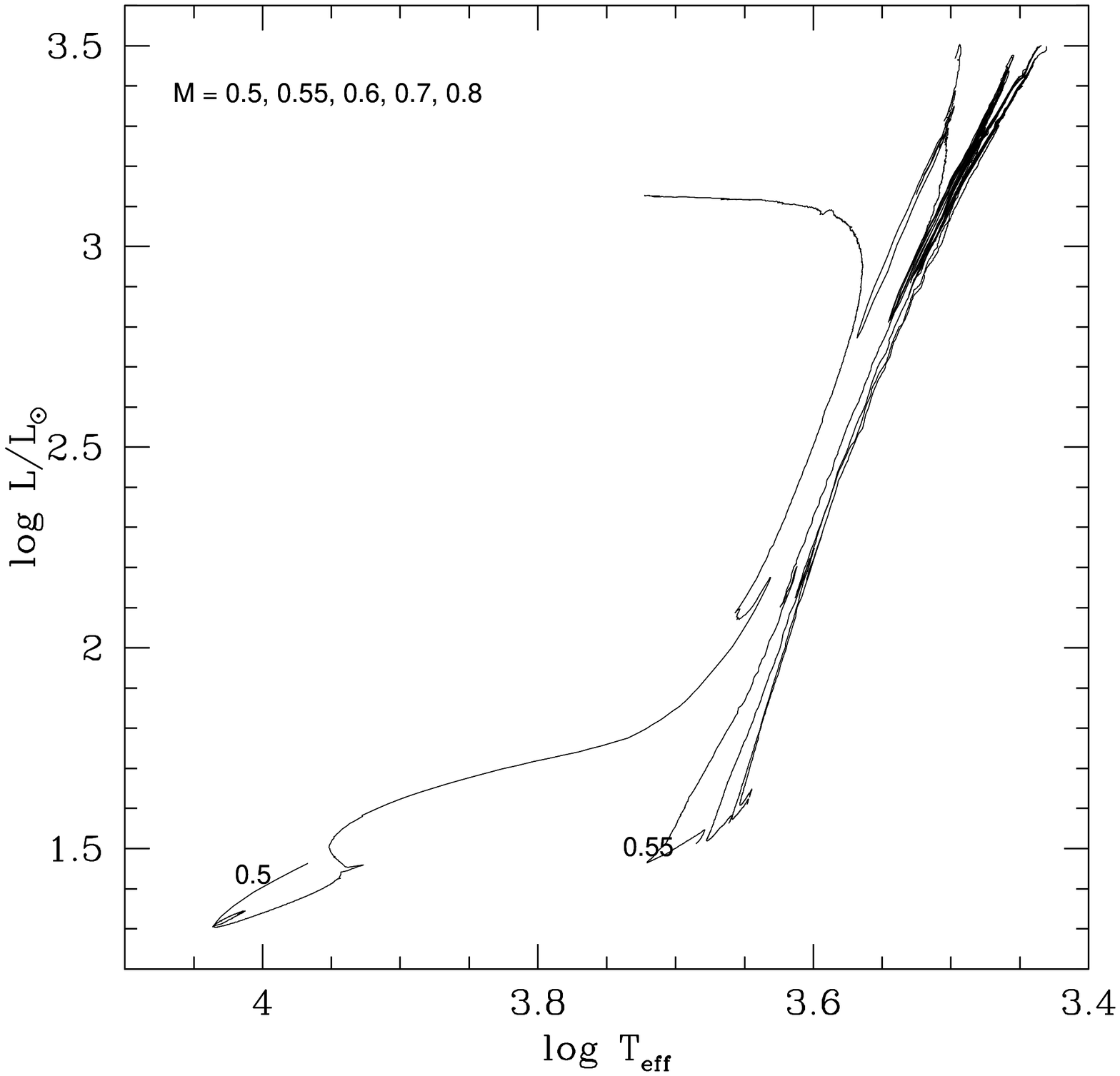}}
\end{minipage} 
\hfill
\begin{minipage}{0.45\textwidth} \noindent d)
\resizebox{\hsize}{!}{\includegraphics{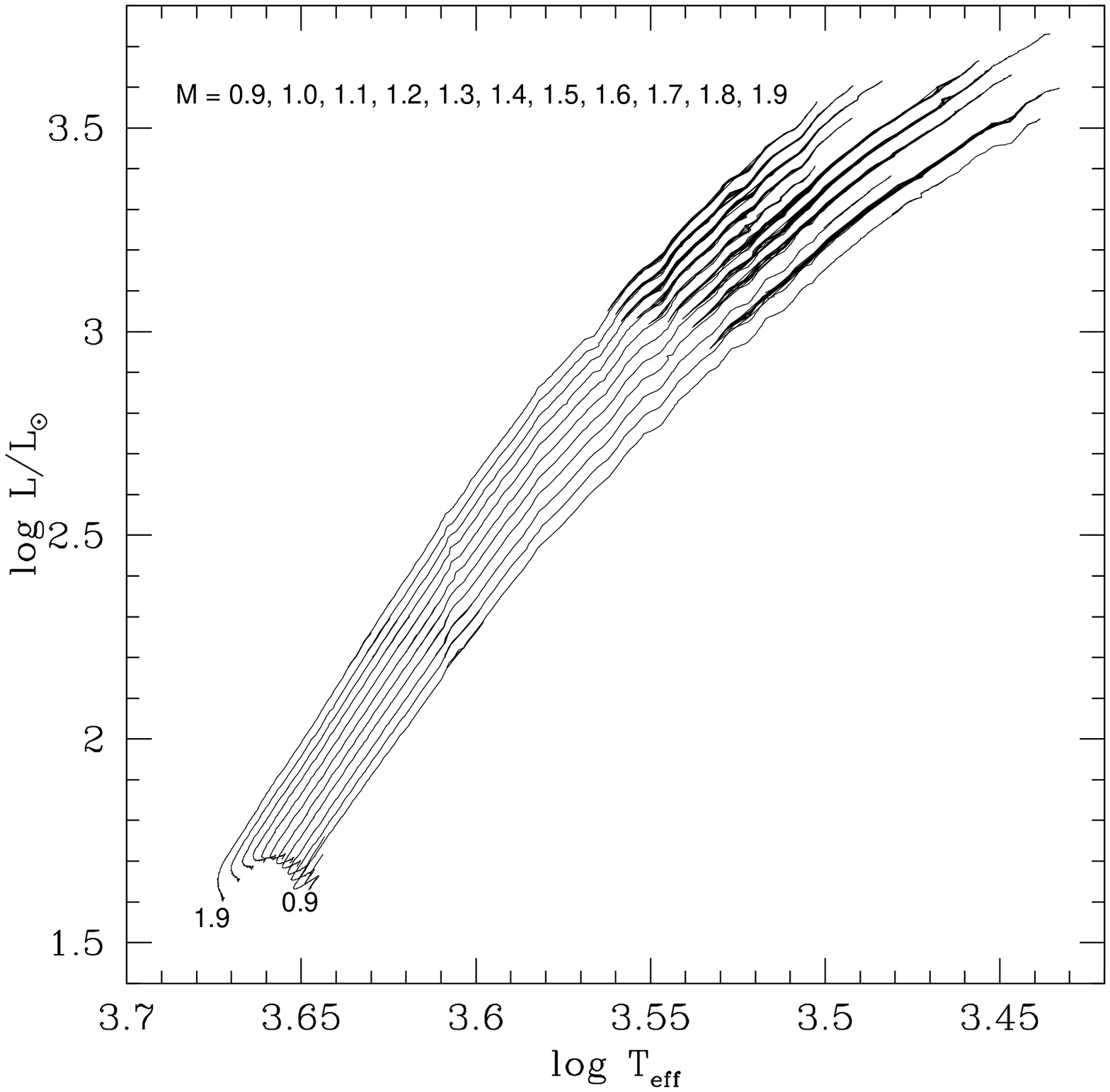}}
\end{minipage} 
\caption{  
The same as \reffig{hrd_z0004}, but for $[Z=0.030, Y=0.300]$.}
\label{hrd_z03}
\end{figure*} 

\section{Stellar tracks}
\label{sec_tracks}

\subsection{Evolutionary stages and mass ranges}
\label{sec_massranges}

Our models are evolved from the ZAMS, at constant mass. 
The evolution through the whole
H- and He-burning phases is followed in detail. The
tracks are stopped either during the TP-AGB phase in intermediate-
and low-mass, or at the onset of carbon ignition in a helium-exhausted
core in the case of our most massive models. In the
case of stellar masses lower than 0.6~\Msun, the main sequence
evolution takes place on time scales much larger than a Hubble time. 
For them, we stopped the computations at an age of about 25~Gyr.

In low-mass stars with $M\ga0.6$~\Msun, the evolution is interrupted
at the stage of He-flash in the electron degenerate hydrogen-exhausted
core. This because the computation of the complete evolution through
the He-flash requires too much CPU time. The evolution is then
re-started from a ZAHB model with the same core mass and surface
chemical composition as the last RGB model. The initial ZAHB model
presents also a core in which 5 percent (in mass fraction) of the
helium has been burned into carbon. This takes into account the
approximate amount of nuclear fuel necessary to lift the core degeneracy
during the He-flash. The evolution is then followed up to the
thermally pulsing AGB phase.

Additional He-burning models with $0.5\le(M/\Msun)<0.6$ have been
computed, starting from a ZAHB model with the same core mass and
surface chemical composition as the last 0.6~\Msun\ RGB model.

In intermediate-mass stars, the evolution goes from the ZAMS up to
either the beginning of the TP-AGB phase, or to the carbon ignition in 
our most massive models (i.e.\ those with masses higher than about
5~\Msun).

For the stellar masses in which the evolution goes through the TP-AGB,
a small number of thermal pulses has been followed -- typically, from
2 to 5 ones. In some few cases, however, the sequences contain only
the first significant pulse, whereas few sequences present as
much as 19 thermal pulse cycles.

Table~\ref{tab_mhef} gives the values of the transition masses 
\Mhef\ and \Mup, as derived from the present tracks. \Mhef\ is 
the maximum mass for a star to develop an electron-degenerate core 
after the main sequence, and sets the limit between low- and 
intermediate-mass stars (see e.g.\ Bertelli et al.\ 1986;
Chiosi et al.\ 1992). For 
deriving the values of Table~\ref{tab_mhef}, we have selected the 
stellar track for which the core mass at He-ignition presents its 
minimum value (see figure 1 in Girardi 1999). It coincides, in 
most cases, with the least massive stellar track we were able to 
evolve through the He-flash. Given the low mass separation between 
the tracks we computed, the \Mhef\ values here presented are 
uncertain by only 0.05~\Msun.

\Mup\ is the maximum mass for a star to develop an
electron-degenerate core after the CHeB phase, and sets the 
limit between intermediate- and high-mass stars (Chiosi et al.\ 
1992). Stars with $M>\Mup$ should ignite carbon and avoid the
thermally pulsing AGB phase. Since we do not follow the carbon
burning in detail, as soon a small amount of carbon burning 
occurs in the stars with mass above $\sim4.5$~\Msun, we 
are not able to determine with confidence whether such burning
will increase or fade away with time (giving place to
an AGB star). Therefore, in Table~\ref{tab_mhef} we simply give
a range of possible values to \Mup, where the lower limit 
represents stars which probably enter in the double-shell 
thermally pulsing phase, and the upper limit that of stars 
which apparently burn carbon explosively. 

\begin{table}
\caption{The transition masses \Mhef\ and \Mup.}
\label{tab_mhef}
\begin{tabular}{lllll}
\noalign{\smallskip}\hline\noalign{\smallskip}
$Z$ & $Y$ & overshoot & $\Mhef/\Msun$ & $\Mup/\Msun$ \\
\noalign{\smallskip}\hline\noalign{\smallskip}
0.0004 & 0.230 & moderate & 1.7 & $4.5-5.0$ \\
0.001 & 0.230 & moderate & 1.7 & $4.5-5.0$ \\
0.004 & 0.240 & moderate & 1.8 & $4.5-5.0$ \\
0.008 & 0.250 & moderate & 1.9 & $4.5-5.0$ \\
0.019 & 0.273 & moderate & 2.0 & $5.0-6.0$ \\
0.030 & 0.300 & moderate & 2.1 & $5.0-6.0$ \\
0.019 & 0.273 & no       & 2.4 & $6.0-7.0$ \\
\noalign{\smallskip}\hline\noalign{\smallskip}
\end{tabular}
\end{table}

\subsection{Tracks in the HR diagram}
\label{sec_hrd}

The complete set of tracks for very low-mass stars ($M<0.6$~\Msun) are 
presented in the HR diagram of \reffig{fig_bassa}. The tracks start at
a stage identified with the ZAMS, and end at the age of 25 Gyr. The ZAMS
model is defined to be the stage of minimum \Teff\ along the computed
track; it follows a stage of much faster evolution in which the pp-cycle 
is out of equilibium, and in which gravitation may provide a
non negligible fraction of the radiated energy.  
It is evident from \reffig{fig_bassa} that these stars evolve very 
little during the Hubble time.

The complete sets of evolutionary tracks with $[Z=0.0004, Y=0.273]$
and $[Z=0.03, Y=0.30]$ are presented in the HR diagrams of 
Figs.~\ref{hrd_z0004} and \ref{hrd_z03}, respectively. In these 
figures, panel (a) presents the low-mass tracks from the 
ZAMS up to the RGB-tip, panel (b) the intermediate-mass ones from the
ZAMS up to the last computed model (either on the TP-AGB phase or 
during C-ignition, depending on the mass), whereas panels (c) and (d)
present the low-mass tracks from the ZAHB up to the TP-AGB phase.  
Figs.~\ref{hrd_z0004} and \ref{hrd_z03} are aimed to illustrate the
typical features and mass coverage of the present tracks, at the 
extreme values of metallicity for which they have been computed.
The reader can notice, for instance, the extended Cepheid
loops present in the $Z=0.0004$ intermediate-mass models, which are 
practically missing in the $Z=0.03$ ones.

We also computed an additional set of ``canonical'' evolutionary tracks
with solar composition, i.e.\ $[Z=0.019, Y=0.273]$. 
They  differ from the models previously described
only in the prescription for the mixing: they are computed assuming the 
classical Schwarzschild criterion for the convective boundaries (i.e.\
without overshooting). Semi-convection is assumed during the core-He 
burning phase.
This set is presented only for $M\ge1.2$~\Msun, since stars with 
$M\la1.1$~\Msun\ present a radiative core during the main sequence; 
therefore, the tracks of lower mass are  
not affected by the adoption of an overshooting scheme
(see also \refsec{sec_conv}), at least in the
main sequence phase. 
The two sets of tracks for $[Z=0.019, Y=0.273]$
provide a useful data-base for comparing 
the behaviour of canonical and overshooting models.

\subsection{Description of the tables}
\label{sec_tabletrack}

The data tables for the present evolutionary tracks are available only in
electronic format. They are stored at the CDS data center in Strasbourg,
and are also available upon request to the authors.  A WWW site with the 
complete data-base (including additional data and the future extensions) 
will be mantained at http://pleiadi.pd.astro.it. 

For each evolutionary track, the corresponding data file presents 
\ref{item_stage} columns with the following information:
	\begin{enumerate}
	\item \verb$age/yr$: stellar age in yr;
	\item \verb$logL$: logarithm of surface luminosity (in solar units), 
\logL;
	\item \verb$logTef$: logarithm of effective temperature (in K), 
\logTe;
	\item \verb$grav$: logarithm of surface gravity ( in cgs units);
	\item \verb$logTc$: logarithm of central temperature (in K);
	\item \verb$logrho$: logarithm of central density (in cgs units);
	\item \verb$Xc,Yc$: mass fraction of either hydrogen (up to the 
central H-exhaustion) or helium (later stages) in the stellar centre;
	\item \verb$Xc_C$: mass fraction of carbon in the stellar centre;
	\item \verb$Xc_O$: mass fraction of oxygen in the stellar centre;
	\item \verb$Q_conv$: fractionary mass of the convective core;
	\item \verb$Q_disc$: fractionary mass of the first mesh point where 
the chemical composition differs from the surface value;
	\item \verb$logL_H$: logarithm of the total luminosity (in solar 
units) provided by H-burning reactions;
	\item \verb$Q1_H$: fractionary mass of the inner border of the 
H-rich region;
	\item \verb$Q2_H$: fractionary mass of the outer border of the 
H-burning region;
	\item \verb$logL_He$: logarithm of the total luminosity (in solar 
units) provided by He-burning reactions; a null value indicates negligible 
energy generation by those reactions;
	\item \verb$Q1_He$: fractionary mass of the inner border of the 
He-burning region;
	\item \verb$Q2_He$: fractionary mass of the outer border of the 
He-burning region;
	\item \verb$logL_C$: logarithm of the total luminosity (in solar 
units) provided by C-burning reactions; a null value means that it is 
negligible;
	\item \verb$logL_nu$: logarithm of the total luminosity (in solar 
units) lost by neutrinos; a null value means that it is negligible;
	\item \verb$Q_Tmax$: fractionary mass of the point with the highest 
temperature inside the star;
	\item \verb$stage$: label indicating particular evolutionary stages.
\label{item_stage}
	\end{enumerate}

A number of evolutionary stages are 
indicated along the tracks (column~\ref{item_stage}). They 
correspond either to: the initial evolutionary stages (\verb$ZAMS$ or 
\verb$ZAHB$), local maxima and minima of $L$ and \Teff\ (\verb$Te-M$, 
\verb$Te-m$, \verb$L-M$, and \verb$L-m$), the exhaustion of central 
hydrogen (\verb$Xc=0$) and helium (\verb$Yc=0$), the 
lowest $L$ and highest \Teff\ during the He-burning of 
intermediate-mass stars (\verb$Bhe$ and \verb$LpT$, respectively), 
the base and tip of the first ascent of the red giant branch (\verb$Brg$ 
and \verb$Tip$, respectively), the maximum $L$ immediately preceding a 
thermal pulse (\verb$1tp$), and the onset of C-burning (\verb$Cb$).
These stages delimit characteristic evolutionary phases, and can be useful 
for the derivation of physical quantities (as e.g.\ typical lifetimes) as a 
function of either mass or metallicity.
Notice that some of these evolutionary stages may be absent from
particular tracks, depending on the precise value of stellar mass 
and metallicity. 

For the sake of conciseness and homogeneity, the evolutionary
tracks in the data-base do not include the fraction of the TP-AGB 
evolution which was actually computed, and which is also presented 
in Figs.~\ref{hrd_z0004} and \ref{hrd_z03}. 
Detailed data about the initial TP-AGB evolution, for any metallicity, 
can be obtained upon request to the authors. However, we remark that
the complete TP-AGB evolution is included, in a synthetic way, in the 
isochrones to be described below.

	\begin{figure}
\resizebox{\hsize}{!}{\includegraphics{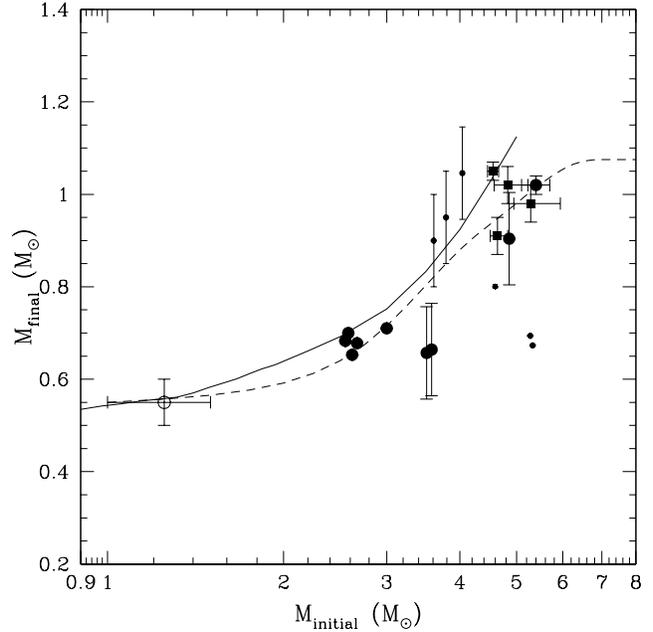}}
	\caption{Comparison between the empirical and theoretical 
initial--final mass relations. The dots are data for WDs in open 
clusters, according to Herwig (1996; full circles) and 
Jeffries (1997, full squares). The smaller dots represent data 
points of lower quality (cf.\ Herwig 1996). The open circle instead
represents the mean masses of field white dwarfs and their 
progenitors. The mean initial--final 
mass relation from Herwig is also shown (dashed line); it is based 
only in the most reliable mass determinations for white dwarfs.
The continuous line, instead, is the initial--final mass relation 
as derived from our $Z=0.019$ models (see text for details).}
	\label{fig_minifin}
	\end{figure}

\begin{figure*}
\begin{minipage}{0.48\textwidth} \noindent a)
\resizebox{\hsize}{!}{\includegraphics{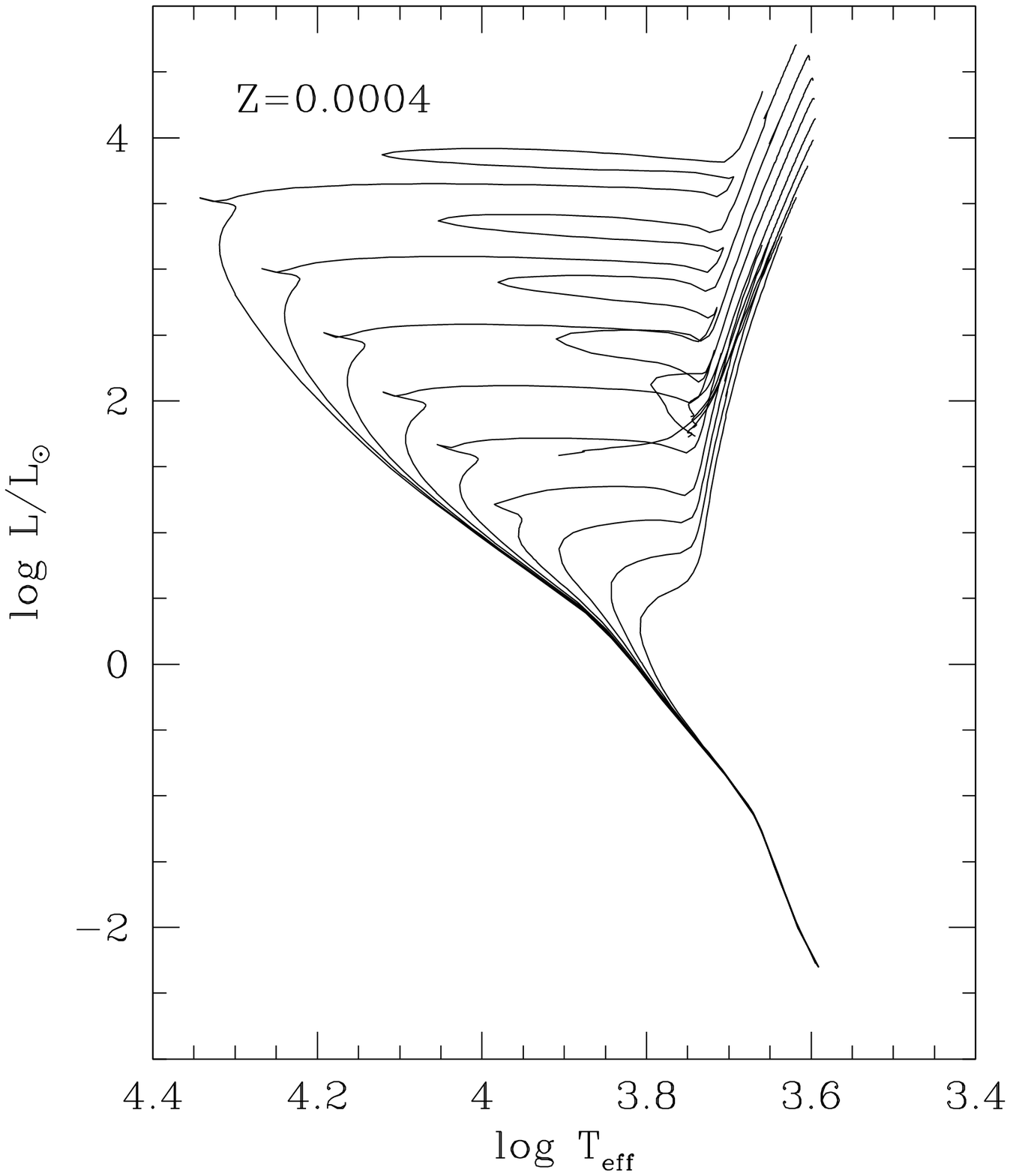}}
\end{minipage} 
\hfill
\begin{minipage}{0.48\textwidth} \noindent b) 
\resizebox{\hsize}{!}{\includegraphics{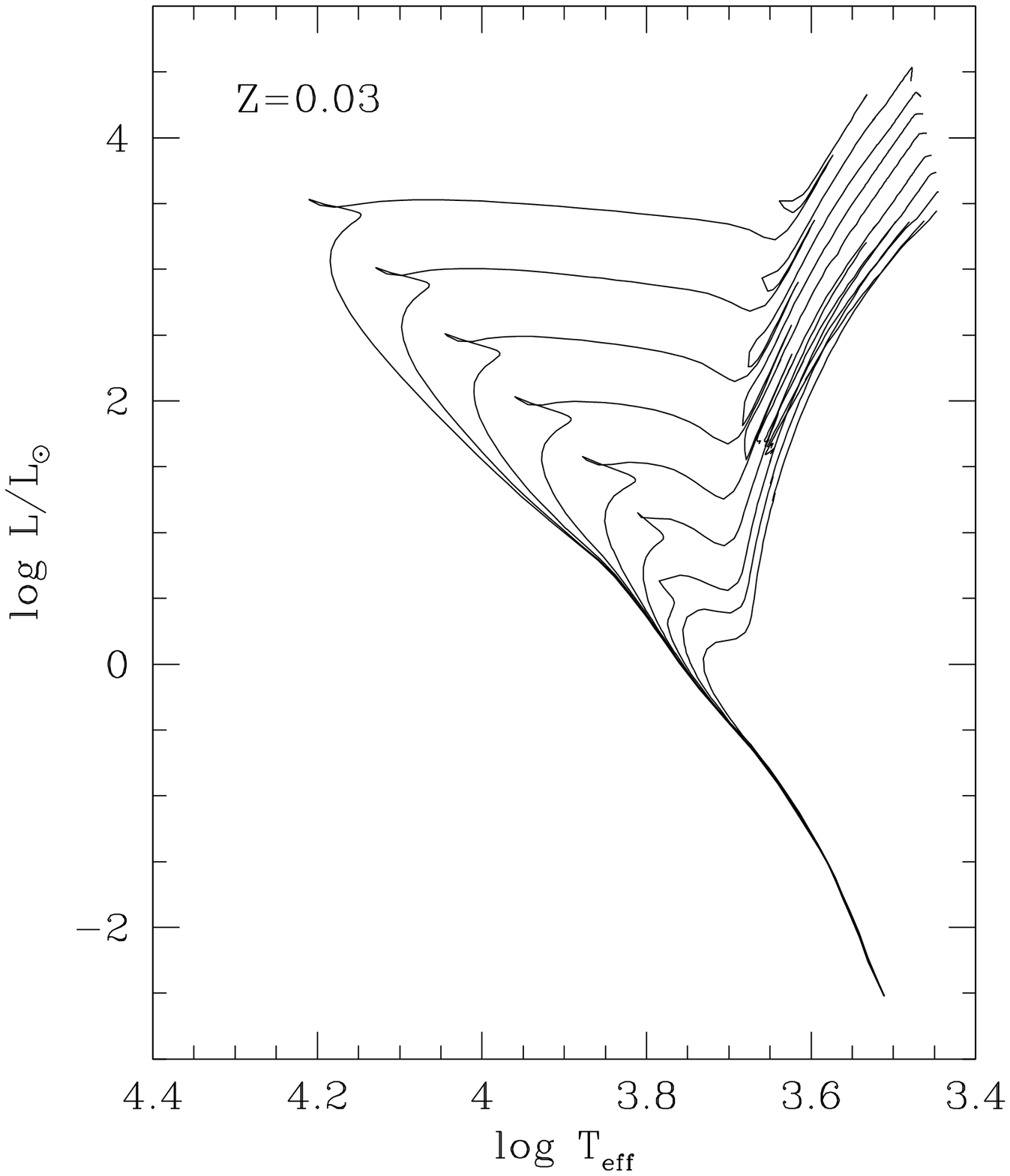}}
\end{minipage} 
\caption{  
Theoretical isochrones in the HR diagram, for the compositions $[Z=0.0004,
Y=0.230]$ (panel a), and $[Z=0.030, Y=0.300]$ (panel b). The age range 
goes from $\logt=7.8$ to 10.2, at equally spaced intervals of
$\Delta\log t=0.3$. In both cases, the main sequence is
complete down to 0.15~\Msun.}
\label{fig_isocrone}
\end{figure*}

\subsection{Changes in surface chemical composition}
\label{sec_chemical}

The surface chemical composition of the stellar models change 
on two well-defined dredge-up events. The first one occurs at
the first ascent of the RGB for all stellar models (except for 
the very-low mass ones which are not evolved out of the main 
sequence). The second dredge-up is found after the core 
He-exhaustion, being remarkable only in models with 
$M\ga3.5$~\Msun. We provide tables with the surface chemical 
composition of H, $^3$He, $^4$He, and main CNO isotopes, before
and after these events (when present in our data). 
Table~\ref{tab_du} shows, as an example, the table for
the solar composition.

\begin{table*}
\caption{Surface chemical composition (by mass fraction) of 
$[Z=0.019, Y=0.273]$ models.}
\label{tab_du}
\begin{tabular}{lllllllllll}
\noalign{\smallskip}\hline\noalign{\smallskip}
$M/\Msun$ &	 H  &    $^3$He  &      $^4$He  &  $^{12}$C   &    $^{13}$C   &    $^{14}$N   &    $^{15}$N  &     $^{16}$O   &    $^{17}$O  &     $^{18}$O \\
\noalign{\smallskip}\hline\noalign{\smallskip}
\multicolumn{11}{l}{Initial:}   \\
  all   & 0.708 & 2.92$\:10^{-5}$ &  0.273 & 3.26$\:10^{-3}$ & 3.92$\:10^{-5}$ & 1.01$\:10^{-3}$ & 3.97$\:10^{-6}$ & 9.15$\:10^{-3}$ & 3.70$\:10^{-6}$ & 2.06$\:10^{-5}$ \\
\noalign{\smallskip}\hline\noalign{\smallskip}
\multicolumn{11}{l}{After the first dredge-up:} \\
  0.60 &  0.697 & 3.03$\:10^{-3}$ &  0.281 & 3.25$\:10^{-3}$ & 4.10$\:10^{-5}$ & 1.01$\:10^{-3}$ & 3.93$\:10^{-6}$ & 9.14$\:10^{-3}$ & 3.71$\:10^{-6}$ & 2.06$\:10^{-5}$ \\
  0.70 &  0.693 & 2.35$\:10^{-3}$ &  0.285 & 3.24$\:10^{-3}$ & 5.62$\:10^{-5}$ & 1.01$\:10^{-3}$ & 3.71$\:10^{-6}$ & 9.15$\:10^{-3}$ & 3.71$\:10^{-6}$ & 2.06$\:10^{-5}$ \\
  0.80 &  0.691 & 1.86$\:10^{-3}$ &  0.288 & 3.16$\:10^{-3}$ & 8.49$\:10^{-5}$ & 1.07$\:10^{-3}$ & 3.47$\:10^{-6}$ & 9.14$\:10^{-3}$ & 3.71$\:10^{-6}$ & 2.05$\:10^{-5}$ \\
  0.90 &  0.690 & 1.54$\:10^{-3}$ &  0.290 & 3.04$\:10^{-3}$ & 9.52$\:10^{-5}$ & 1.20$\:10^{-3}$ & 3.27$\:10^{-6}$ & 9.15$\:10^{-3}$ & 3.72$\:10^{-6}$ & 2.03$\:10^{-5}$ \\
  1.00 &  0.690 & 1.20$\:10^{-3}$ &  0.290 & 2.94$\:10^{-3}$ & 9.89$\:10^{-5}$ & 1.32$\:10^{-3}$ & 3.12$\:10^{-6}$ & 9.15$\:10^{-3}$ & 3.75$\:10^{-6}$ & 1.99$\:10^{-5}$ \\
  1.10 &  0.690 & 9.83$\:10^{-4}$ &  0.290 & 2.83$\:10^{-3}$ & 1.00$\:10^{-4}$ & 1.44$\:10^{-3}$ & 2.98$\:10^{-6}$ & 9.15$\:10^{-3}$ & 3.89$\:10^{-6}$ & 1.94$\:10^{-5}$ \\
  1.20 &  0.693 & 8.32$\:10^{-4}$ &  0.287 & 2.77$\:10^{-3}$ & 1.02$\:10^{-4}$ & 1.51$\:10^{-3}$ & 2.89$\:10^{-6}$ & 9.14$\:10^{-3}$ & 4.06$\:10^{-6}$ & 1.91$\:10^{-5}$ \\
  1.30 &  0.695 & 7.40$\:10^{-4}$ &  0.286 & 2.72$\:10^{-3}$ & 9.63$\:10^{-5}$ & 1.57$\:10^{-3}$ & 2.88$\:10^{-6}$ & 9.14$\:10^{-3}$ & 4.31$\:10^{-6}$ & 1.87$\:10^{-5}$ \\
  1.40 &  0.695 & 6.78$\:10^{-4}$ &  0.286 & 2.61$\:10^{-3}$ & 1.01$\:10^{-4}$ & 1.70$\:10^{-3}$ & 2.68$\:10^{-6}$ & 9.14$\:10^{-3}$ & 4.86$\:10^{-6}$ & 1.82$\:10^{-5}$ \\
  1.50 &  0.693 & 5.92$\:10^{-4}$ &  0.288 & 2.39$\:10^{-3}$ & 1.07$\:10^{-4}$ & 1.95$\:10^{-3}$ & 2.38$\:10^{-6}$ & 9.14$\:10^{-3}$ & 6.47$\:10^{-6}$ & 1.72$\:10^{-5}$ \\
  1.60 &  0.693 & 5.22$\:10^{-4}$ &  0.287 & 2.33$\:10^{-3}$ & 1.06$\:10^{-4}$ & 2.03$\:10^{-3}$ & 2.31$\:10^{-6}$ & 9.12$\:10^{-3}$ & 2.00$\:10^{-5}$ & 1.68$\:10^{-5}$ \\
  1.70 &  0.693 & 4.57$\:10^{-4}$ &  0.288 & 2.28$\:10^{-3}$ & 1.05$\:10^{-4}$ & 2.12$\:10^{-3}$ & 2.26$\:10^{-6}$ & 9.06$\:10^{-3}$ & 3.80$\:10^{-5}$ & 1.65$\:10^{-5}$ \\
  1.80 &  0.692 & 4.08$\:10^{-4}$ &  0.288 & 2.29$\:10^{-3}$ & 1.08$\:10^{-4}$ & 2.16$\:10^{-3}$ & 2.24$\:10^{-6}$ & 9.00$\:10^{-3}$ & 4.28$\:10^{-5}$ & 1.64$\:10^{-5}$ \\
  1.90 &  0.691 & 3.65$\:10^{-4}$ &  0.289 & 2.25$\:10^{-3}$ & 1.07$\:10^{-4}$ & 2.25$\:10^{-3}$ & 2.20$\:10^{-6}$ & 8.93$\:10^{-3}$ & 5.13$\:10^{-5}$ & 1.62$\:10^{-5}$ \\
  1.95 &  0.691 & 3.45$\:10^{-4}$ &  0.290 & 2.25$\:10^{-3}$ & 1.07$\:10^{-4}$ & 2.28$\:10^{-3}$ & 2.20$\:10^{-6}$ & 8.91$\:10^{-3}$ & 4.50$\:10^{-5}$ & 1.62$\:10^{-5}$ \\
  2.00 &  0.690 & 3.25$\:10^{-4}$ &  0.291 & 2.22$\:10^{-3}$ & 1.03$\:10^{-4}$ & 2.35$\:10^{-3}$ & 2.19$\:10^{-6}$ & 8.86$\:10^{-3}$ & 5.27$\:10^{-5}$ & 1.61$\:10^{-5}$ \\
  2.20 &  0.687 & 2.61$\:10^{-4}$ &  0.294 & 2.20$\:10^{-3}$ & 1.06$\:10^{-4}$ & 2.48$\:10^{-3}$ & 2.17$\:10^{-6}$ & 8.74$\:10^{-3}$ & 5.13$\:10^{-5}$ & 1.59$\:10^{-5}$ \\
  2.50 &  0.683 & 2.09$\:10^{-4}$ &  0.298 & 2.21$\:10^{-3}$ & 1.08$\:10^{-4}$ & 2.62$\:10^{-3}$ & 2.13$\:10^{-6}$ & 8.58$\:10^{-3}$ & 4.93$\:10^{-5}$ & 1.59$\:10^{-5}$ \\
  3.00 &  0.680 & 1.46$\:10^{-4}$ &  0.301 & 2.19$\:10^{-3}$ & 1.08$\:10^{-4}$ & 2.76$\:10^{-3}$ & 2.09$\:10^{-6}$ & 8.47$\:10^{-3}$ & 3.08$\:10^{-5}$ & 1.58$\:10^{-5}$ \\
  3.50 &  0.682 & 1.11$\:10^{-4}$ &  0.299 & 2.23$\:10^{-3}$ & 1.08$\:10^{-4}$ & 2.71$\:10^{-3}$ & 2.12$\:10^{-6}$ & 8.48$\:10^{-3}$ & 2.13$\:10^{-5}$ & 1.60$\:10^{-5}$ \\
  4.00 &  0.682 & 9.04$\:10^{-5}$ &  0.299 & 2.26$\:10^{-3}$ & 1.15$\:10^{-4}$ & 2.67$\:10^{-3}$ & 2.11$\:10^{-6}$ & 8.48$\:10^{-3}$ & 1.49$\:10^{-5}$ & 1.63$\:10^{-5}$ \\
  4.50 &  0.681 & 7.52$\:10^{-5}$ &  0.299 & 2.27$\:10^{-3}$ & 1.17$\:10^{-4}$ & 2.69$\:10^{-3}$ & 2.10$\:10^{-6}$ & 8.44$\:10^{-3}$ & 1.19$\:10^{-5}$ & 1.64$\:10^{-5}$ \\
  5.00 &  0.680 & 6.37$\:10^{-5}$ &  0.300 & 2.30$\:10^{-3}$ & 1.16$\:10^{-4}$ & 2.70$\:10^{-3}$ & 2.12$\:10^{-6}$ & 8.39$\:10^{-3}$ & 1.16$\:10^{-5}$ & 1.65$\:10^{-5}$ \\
  6.00 &  0.681 & 4.93$\:10^{-5}$ &  0.300 & 2.35$\:10^{-3}$ & 1.21$\:10^{-4}$ & 2.66$\:10^{-3}$ & 2.14$\:10^{-6}$ & 8.36$\:10^{-3}$ & 8.92$\:10^{-6}$ & 1.67$\:10^{-5}$ \\
  7.00 &  0.679 & 4.03$\:10^{-5}$ &  0.302 & 2.33$\:10^{-3}$ & 1.20$\:10^{-4}$ & 2.76$\:10^{-3}$ & 2.13$\:10^{-6}$ & 8.28$\:10^{-3}$ & 7.84$\:10^{-6}$ & 1.66$\:10^{-5}$ \\
\noalign{\smallskip}\hline\noalign{\smallskip}
\multicolumn{11}{l}{After the second dredge-up:} \\
  4.50 &  0.653 & 7.06$\:10^{-5}$ &  0.328 & 2.14$\:10^{-3}$ & 1.14$\:10^{-4}$ & 3.15$\:10^{-3}$ & 1.98$\:10^{-6}$ & 8.09$\:10^{-3}$ & 1.30$\:10^{-5}$ & 1.55$\:10^{-5}$ \\
  5.00 &  0.635 & 5.80$\:10^{-5}$ &  0.346 & 2.11$\:10^{-3}$ & 1.13$\:10^{-4}$ & 3.42$\:10^{-3}$ & 1.94$\:10^{-6}$ & 7.82$\:10^{-3}$ & 1.23$\:10^{-5}$ & 1.51$\:10^{-5}$ \\
  6.00 &  0.610 & 4.29$\:10^{-5}$ &  0.371 & 2.07$\:10^{-3}$ & 1.16$\:10^{-4}$ & 3.74$\:10^{-3}$ & 1.89$\:10^{-6}$ & 7.51$\:10^{-3}$ & 9.47$\:10^{-6}$ & 1.49$\:10^{-5}$ \\
  7.00 &  0.622 & 3.56$\:10^{-5}$ &  0.359 & 2.09$\:10^{-3}$ & 1.16$\:10^{-4}$ & 3.65$\:10^{-3}$ & 1.90$\:10^{-6}$ & 7.58$\:10^{-3}$ & 8.57$\:10^{-6}$ & 1.48$\:10^{-5}$ \\
\noalign{\smallskip}\hline\noalign{\smallskip}
\end{tabular}
\end{table*}

\section{Mass loss on the RGB, and synthetic TP-AGB evolution}
\label{sec_rgbagb}

Before presenting the isochrones derived from our
evolutionary tracks, we briefly describe the
way we have considered the effect of mass loss on the RGB. 
Also, we describe the extension of our tracks of low- 
and intermediate-mass stars through 
the complete TP-AGB phase. The latter point is an important one,
since this evolutionary phase constitutes a significant
fraction of the bolometric luminosity of stellar populations.

Mass loss by stellar winds during the RGB of low-mass stars 
is considered only at the stage of isochrone construction.
We use the empirical formulation by Reimers (1975), but
with mass-loss rates multiplied by a parameter $\eta$ which is set 
equal to 0.4 (see Renzini \& Fusi Pecci 1988). The procedure is 
basically the following:
In passing from the RGB-tip to the ZAHB, we first integrate the mass 
loss rate along the RGB of every single track, in order
to estimate the total amount of mass that has to be
removed. Then the mass of the evolutionary models (that were computed at
constant mass) is simply scaled down to the value suited to the ZAHB stars.
This approximation is a good one since the mass loss does not 
affect significantly the internal structure of models at the tip of the
RGB.

In addition, before constructing the isochrones, we need to complete 
the stellar evolution along the TP-AGB. This phase is followed 
in a synthetic way (see Iben \& Truran 1978; Groenewegen \& de Jong
1993; Bertelli et al.\ 1994; Marigo et al.\ 1996, 1998;
Girardi \& Bertelli 1998). In a few words, we evolve the core mass,
total mass, effective temperature and luminosity of each star,
from the first thermal pulse on the AGB up to the stage of complete 
envelope ejection. The following relations are used to this aim:
	\begin{itemize}
	\item The core mass--luminosity relation described in 
eqs.\ 10 and 11 of Groenewegen \& de Jong (1993, 
and references therein);
	\item The rate of core growth from 
eq.\ 18 of Groenewegen \& de Jong (1993).
	\item The luminosity--effective temperature relations 
described in eq.\ 12 of Girardi \& Bertelli (1998). Suffice it 
to recall that these relations are obtained by 
extrapolating the slope of the early-AGB phase of our models
to higher luminosities and lower effective temperatures.
This is performed separately for each value of the metallicity.
	\item We adopt the Vassiliadis \& Wood's
(1993) prescription for the mass-loss along the TP-AGB, since it 
proved to provide a reasonable description for the initial--final 
relation of low- and intermediate-mass stars of solar metallicity 
(see e.g.\ Marigo 1998a), as well as to the maximum luminosity of 
AGB stars of different ages (and hence initial masses) observed in 
the LMC (see Marigo et al.\ 1996).
	\end{itemize}

It is worth remarking that this prescription for the TP-AGB 
evolution can be considered only as a crude first approximation to
the real evolution. We do not consider, for instance, key processes
as the third dredge-up and hot-bottom burning in TP-AGB stars. 
We intend to replace soon the present prescription, 
for the detailed TP-AGB models of Marigo et al.\ (1998, 1999).

For the moment, we just compare the initial--final mass relation,
as derived from the present tracks, with the empirical one of 
Herwig (1996). This is done in \reffig{fig_minifin}.
It is important to recall some aspects of the empirical relation. 
First, the Herwig (1996) relation is very different from
the (largely used) Weidemann (1987) one, at least in the range of
initial masses $M\ga2$~\Msun; this is due, essentially, to the 
dramatic re-evaluation of the mass of the white dwarfs in the
Hyades (cf.\ Weidemann 1996). Second, the Herwig 
(1996) relation extends up to a initial mass of 8~\Msun, value 
which represents the white dwarfs in the open cluster NGC~2516. 
Recently, the initial mass of the white dwarfs in this cluster
have been re-evaluated, to new values of $5-6$~\Msun, 
by Jeffries (1997). Therefore, the upper mass limit of stars 
which produce white dwarfs, is roughly consistent with the
values of $\Mup\sim5$~\Msun\ we find in our stellar models.

Figure~\ref{fig_minifin} evidences that our theoretical 
initial--final mass relation for the solar metallicity,
reproduces in a satisfactory way the empirical one from
Herwig (1996) and Jeffries (1997). This is an important point
for a set of isochrones aimed to be used in evolutionary 
population synthesis calculations.

\section{Isochrones}
\label{sec_isochrones}

\subsection{Construction of isochrones}
\label{sec_isochrd}

From the tracks presented in this paper, we have constructed isochrones
adopting the same algorithm of ``equivalent evolutionary points'' 
as used in Bertelli et al.\ (1990, 1994). 

The initial point of each isochrone is the 0.15~\Msun\ model in the 
lower main sequence. The terminal stage of the isochrones is either 
the tip of the TP-AGB for $M\la5$~\Msun (ages of $\sim10^8$~yr), or 
C-ignition in the core for the remaining stars.

Theoretical luminosities and effective temperatures along the isochrones 
are translated to magnitudes and colors using extensive tabulations of 
bolometric corrections and colors, as in Bertelli et al.\ (1994). 
The tabulations were obtained from convolving the spectral 
energy distributions contained in the library of stellar spectra of 
Kurucz (1992) with the response function of several broad-band filters. 
The response functions are from Buser \& Kurucz 
(1978) for the $UBV$ pass-bands, from Bessell (1990) for the $R$ and $I$ 
Cousins, and finally from Bessell \& Brett (1988) for the $JHK$ ones.

\subsection{Description of isochrone tables}
\label{sec_tableisoc}

In Fig.~\ref{fig_isocrone} we present some of the derived isochrones 
on the HRD, sampled at age intervals of $\Delta\log t=0.3$. 
They cover the complete age range from about 0.06 to 16~Gyr. 
Younger isochrones could be constructed only with the aid of 
evolutionary tracks for stars with initial masses $M\ga7$~\Msun, which 
are not presented in this paper.  
                        
Complete tables with the isochrones can be obtaind at the CDS in 
Strasbourg, or upon request to the authors, or through the WWW site 
http://pleiadi.pd.astro.it. In this data-base, isochrones are provided
at $\Delta\log t=0.05$ intervals; this means that any two consecutive
isochrones differ by only 12 percent in their ages.

For each isochrone table, the layout is as follows: 
An header presents the basic information about the age and metallicity
of each isochrone. Column~1 presents the logarithm of the age in yr;
Columns~2 and 3 the initial and actual stellar masses, 
respectively. We recall that the initial mass is the useful quantity
for population synthesis calculations, since together with the initial 
mass function it determines the relative number of stars in different
sections of the isochrones. Then follow the logaritms of surface 
luminosity (column 4), effective temperature (column 5), and surface 
gravity (column 6). From columns 7 to 15, we have the sequence of 
absolute magnitudes, starting with the bolometric one and following 
those in the $UBVRIJHK$ pass-bands. In the last column (16), the 
indefinite integral over the initial mass $M$ of the initial mass 
function (IMF) by number, i.e.\ 
	\begin{equation}
\mbox{\sc flum} = \int\phi(M) \diff M
	\end{equation}
is presented, for the case of the Salpeter IMF, $\phi(M)=AM^{-\alpha}$, 
with $\alpha=2.35$. When we assume a normalization constant of $A=1$, 
{\sc flum} is simply given by {\sc flum}$ = M^{1-\alpha}/(1-\alpha)$.
This is a useful quantity since the difference between any two 
values of {\sc flum} is proportional to the number of stars located in 
the corresponding mass interval. It is worth remarking that we 
present {\sc flum} values 
for the complete mass interval down to 0.15~\Msun, always assuming a
Salpeter IMF, whereas we know that such an IMF cannot be extended to
such low values of the mass. However, the reader can easily derive 
{\sc flum} relations for alternative choices of the IMF, by using 
the values of the initial mass we present in the Column 2 of the
isochrone tables.

\begin{table*}
\caption{Sample summary table with the most significant stages 
along some $Z=0.019$ isochrones.}
\label{tab_iso}
\begin{tabular}{cccccrrrrrl}
\noalign{\smallskip}\hline\noalign{\smallskip}
 $\log({\rm age/yr})$ & $M_{\rm ini}$ & $M_{\rm act}$ & 
\logL & \logTe & $\log G$ & \mv & $U-B$ & $B-V$ & $V-I$ & stage \\
\noalign{\smallskip}\hline\noalign{\smallskip}
  7.80 & 5.6503 & 5.644 & 3.147 & 4.206 & 3.81 & $-$1.725 & $-$0.643 & $-$0.174 & $-$0.161 &  TO  \\     
  7.80 & 6.3176 & 6.305 & 3.474 & 4.164 & 3.37 & $-$2.789 & $-$0.586 & $-$0.156 & $-$0.135 &  Te-m \\      
  7.80 & 6.3592 & 6.346 & 3.572 & 4.223 & 3.51 & $-$2.690 & $-$0.688 & $-$0.187 & $-$0.167 &  Te-M  \\     
  7.80 & 6.3625 & 6.349 & 3.576 & 4.067 & 2.88 & $-$3.584 & $-$0.394 & $-$0.113 & $-$0.068 &  L-M  \\      
  7.80 & 6.3666 & 6.353 & 3.314 & 3.656 & 1.50 & $-$3.026 & 1.170 & 1.222 & 1.155 &  RGBb  \\     
  7.80 & 6.3733 & 6.356 & 3.898 & 3.588 & 0.64 & $-$3.845 & 1.806 & 1.542 & 1.657 &  RGBt  \\     
  7.80 & 6.4397 & 6.397 & 3.491 & 3.635 & 1.24 & $-$3.291 & 1.458 & 1.348 & 1.282 &  BHeb  \\     
  7.80 & 6.4899 & 6.440 & 3.686 & 3.819 & 1.78 & $-$4.523 & 0.230 & 0.363 & 0.431 &  Te-M  \\     
  7.80 & 6.6046 & 6.530 & 3.703 & 3.612 & 0.94 & $-$3.599 & 1.697 & 1.473 & 1.438 &  EHeb  \\     
  7.80 & 6.6152 & 6.526 & 4.317 & 3.553 & 0.09 & $-$4.262 & 1.813 & 1.581 & 2.230 &  Cb    \\     
$\vdots$ & $\vdots$ & $\vdots$ & $\vdots$ & $\vdots$ & $\vdots$ & $\vdots$ & $\vdots$ & $\vdots$ & $\vdots$ & $\vdots$ \\
  9.00 & 1.7933 & 1.791 & 1.160 & 3.871 & 3.96 & 1.826 & 0.025 & 0.270 & 0.308 &  TO     \\    
  9.00 & 2.0510 & 2.047 & 1.443 & 3.825 & 3.55 & 1.148 & 0.004 & 0.414 & 0.495 &  Te-m   \\    
  9.00 & 2.0663 & 2.062 & 1.609 & 3.890 & 3.65 & 0.693 & 0.107 & 0.181 & 0.205 &  Te-M   \\    
  9.00 & 2.0671 & 2.063 & 1.560 & 3.827 & 3.45 & 0.851 & 0.014 & 0.406 & 0.486 &  L-M   \\     
  9.00 & 2.0728 & 2.068 & 1.292 & 3.713 & 3.26 & 1.747 & 0.494 & 0.883 & 0.906 &  RGBb   \\    
  9.00 & 2.0866 & 2.080 & 2.353 & 3.635 & 1.89 & $-$0.454 & 1.293 & 1.283 & 1.280 &  RGBt  \\     
  9.00 & 2.0913 & 2.084 & 1.528 & 3.691 & 2.94 & 1.253 & 0.683 & 0.991 & 0.996 &  BHeb   \\    
  9.00 & 2.1686 & 2.159 & 1.620 & 3.692 & 2.87 & 1.016 & 0.677 & 0.987 & 0.989 &  Te-M   \\    
  9.00 & 2.2986 & 2.283 & 2.113 & 3.664 & 2.28 & $-$0.056 & 0.978 & 1.135 & 1.124 &  EHeb  \\     
  9.00 & 2.3080 & 2.279 & 3.326 & 3.564 & 0.67 & $-$2.051 & 1.818 & 1.570 & 2.000 &  1TP   \\     
  9.00 & 2.3096 & 0.672 & 4.069 & 3.473 & $-$0.97 & $-$1.287 & 1.377 & 1.617 & 3.764 &  AGBt \\ 
$\vdots$ & $\vdots$ & $\vdots$ & $\vdots$ & $\vdots$ & $\vdots$ & $\vdots$ & $\vdots$ & $\vdots$ & $\vdots$ & $\vdots$ \\
  10.20 & 0.8938 & 0.892 & 0.090 & 3.742 & 4.21 & 4.665 & 0.269 & 0.748 & 0.791 &  TO    \\     
  10.20 & 0.9190 & 0.917 & 0.240 & 3.702 & 3.91 & 4.436 & 0.557 & 0.917 & 0.955 &  RGBb   \\    
  10.20 & 0.9321 & 0.927 & 1.350 & 3.653 & 2.61 & 1.928 & 1.021 & 1.162 & 1.184 &  L-M   \\     
  10.20 & 0.9324 & 0.926 & 1.289 & 3.656 & 2.69 & 2.057 & 0.984 & 1.143 & 1.167 &  L-m   \\     
  10.20 & 0.9342 & 0.743 & 3.370 & 3.485 & $-$0.18 & 0.191 & 1.399 & 1.615 & 3.577 &  RGBt  \\     
  10.20 & 0.9342 & 0.743 & 1.574 & 3.666 & 2.35 & 1.276 & 0.948 & 1.121 & 1.112 &  BHeb    \\   
  10.20 & 0.9343 & 0.742 & 1.625 & 3.661 & 2.27 & 1.180 & 0.999 & 1.146 & 1.135 &  Te-m   \\    
  10.20 & 0.9354 & 0.738 & 1.586 & 3.670 & 2.35 & 1.220 & 0.916 & 1.104 & 1.093 &  Te-M  \\     
  10.20 & 0.9363 & 0.735 & 2.019 & 3.631 & 1.76 & 0.415 & 1.360 & 1.311 & 1.303 &  EHeb  \\     
  10.20 & 0.9363 & 0.734 & 2.267 & 3.609 & 1.42 & 0.017 & 1.647 & 1.450 & 1.454 &  L-M  \\      
  10.20 & 0.9363 & 0.733 & 2.142 & 3.620 & 1.59 & 0.210 & 1.502 & 1.377 & 1.373 &  L-m  \\      
  10.20 & 0.9365 & 0.674 & 3.314 & 3.487 & $-$0.15 & 0.275 & 1.404 & 1.614 & 3.537 &  1TP  \\      
  10.20 & 0.9365 & 0.529 & 3.499 & 3.465 & $-$0.53 & 0.315 & 1.362 & 1.618 & 3.886 &  AGBt \\ 
\noalign{\smallskip}\hline\noalign{\smallskip}
\end{tabular}
\end{table*}

We also provide summary tables containing
basic information for the most significant stages along the 
isochrones. A sample table of this kind is presented in 
\reftab{tab_iso} below, for three $Z=0.019$ isochrones, with age
values $\log({\rm age/yr})=7.8$, 9.0 and 10.2. 
The evolutionary stages are listed in the last column, and are, 
in sequence:
	\begin{itemize}
	\item \verb$TO$: the turn-off point, i.e.\ 
the point of highest \Teff\ during the core$-$H burning phase.
	\item If present, \verb$Te-m$ and \verb$Te-M$ signal 
the coldest and hottest points, respectively, of stars in the
overall contraction phase at the end of core$-$H burning; 
in this case \verb$Te-M$ roughly corresponds to the stars in the
stage of core H$-$exhaustion. Occasionally, this stage is followed by 
a local maximum of luminosity, \verb$L-M$, of stars which are crossing 
Hertzsprung gap.
	\item \verb$RGBb$: the base of the RGB.
	\item if present, \verb$L-M$ and \verb$L-m$ limit
the luminosity interval of RGB stars which are crossing the 
discontinuity in chemical profile left by the first dredge$-$up 
event; so, this interval corresponds to the bump in the
luminosity function along the RGB.
	\item \verb$RGBt$: the tip of the RGB.
	\item \verb$BHeb$: the beginning of the CHeB phase. It is 
defined as the point of lowest luminosity for CHeB stars.
	\item If present, \verb$Te-m$ and \verb$Te-M$ signal 
the coldest and hottest points, respectively, for CHeB stars.
For the youngest isochrones, \verb$Te-M$ represents the maximum 
extension of the Cepheid loop.
	\item \verb$EHeb$: the end of the CHeB phase.
	\item In the oldest isochrones, \verb$L-M$ and \verb$L-m$ 
limit the luminosity range of early-AGB stars; this interval 
corresponds to the clump of early-AGB stars in colour-magnitude 
diagrams.
	\item \verb$1TP$: the beginning of the thermally pulsing 
AGB phase. 
	\item \verb$AGBt$: the end of the AGB phase.
	\item \verb$Cb$: the stage of C-ignition in the cases the 
AGB phase do not occur.
	\end{itemize}
Similar tables are presented in the data-base, for other values of 
age and metallicity.

In addition, we provide tables with the integrated broad-band colours 
of single-burst stellar populations. Such tables are derived
by integrating the stellar luminosities, weighted by the predicted
number of stars in each bin, along the isochrones.

\section{Concluding remarks}
\label{sec_remarks}

The stellar evolutionary tracks described here constitute a very
extensive and homogeneous grid. They are therefore suited to the
purposes of evolutionary population synthesis, either of simple (star
clusters) or complex (galaxies) stellar populations.  They can be 
used to describe populations older than $6\times10^7$~yr, for a wide 
range of metallicities. This range does not include, however, that of 
very low-metallicities found in some globular clusters, and
the tail of very high 
metallicities which may be found in the most massive elliptical 
galaxies and bulges. We intend to extend the data-base in order to 
cover these intervals in forthcoming papers.

One of the main characteristic of this data-base is that stellar
tracks are presented at very small mass intervals. The typical mass
resolution for low-mass stars is of 0.1~\Msun. This is reduced to
0.05~\Msun\ in the interval of very-low masses ($M\la0.6$~\Msun), and
occasionally in the vicinity of the limit mass \Mhef\ between low- and
intermediate mass stars. The mass separation between tracks
increases in the range of intermediate-mass stars, 
but anyway we provide enough tracks
to allow a very detailed mapping of the HR diagram.
As a result of this good mass resolution, the derived theoretical 
isochrones are also very detailed.

Due to their characteristics, the data-base is particularly suited to
the construction of synthetic colour-magnitude diagrams
(CMD). The latter are important tools for the correct interpretation
of the high-quality photometric data which is becoming available for
Local Group galaxies.  Among these data, we mention the HIPPARCOS
results for the solar vicinity (Perryman et al.\ 1995), the extensive 
photometric data-bases
derived from the search of micro-lensing events towards
the Magellanic Clouds and the Galactic Bulge, and deep HST images of
particular galaxy fields.

As examples of the level of detail provided by the present tracks and 
isochrones, we mention the results obtained in the works by Girardi 
et al.\ (1998) and Girardi (1999). They have simulated 
synthetic CMDs for different model galaxies, finding substructures in the
red clump region of the CMD. In particular, a faint secondary clump
happens to be present is such models, as a result of the fine mass 
resolution adopted in the calculation of the evolutionary tracks
with mass $M\sim2$~\Msun. 
Conterparts to these clump substructures have
been noticed in the CMDs derived from HIPPARCOS data, and those of 
some Magellanic Cloud fields observed to date (see Girardi 1999). 

In addition, the present tracks behave in a very regular way for
different metallicities, so that the construction of isochrones for
any intermediate value of metallicity (in the interval $0.0004\le
Z\le0.03$) is possible. These isochrones for arbitrary values of $Z$ 
are available upon request, but are not included
in the present electronic data-base.

An open question is whether the present models can be complemented
with those of Bertelli et al.\ (1994) and Girardi et al.\ (1996a), for
metallicities lower than $Z=0.0004$ and higher than $Z=0.03$, or masses
higher than 7~\Msun. The answer is not simple. In fact there are
several differences between both sets of models. First of all, for
metallicities higher than $Z=0.008$, the adopted $Y(Z)$ relations are
different. It generates systematic, although small, differences in the
model luminosities and lifetimes for the sets of highest
metallicities. Whether they are significant, however, is something 
that depends on the level of detail one is interested to look at.  

The Bertelli et al.\ (1994) data-base does not contain $Z=0.001$ 
models with OPAL opacities, which are included in the present one. 
With respect to lower metallicities, the Girardi et al.\ (1996a) models
with $Z=0.0001$, are probably a valid complement to the present ones, but 
present a lower mass resolution than now adopted, and a different 
prescription for overshooting in the mass range $1.0<(M/\Msun)<1.5$.  

On the other hand, a preliminary comparison between the present and 
previous Padova evolutionary models, in the range from about 5 to 7~\Msun, 
reveals that they have almost identical tracks in the HRD and lifetimes. 
Therefore, they can be used to complement the present models for masses
higher than 7~\Msun, or equivalently for ages younger than $10^8$~yr.


\subsection*{Acknowledgements}

We thank D.\ Mihalas and R.\ Wehrse for kindly providing us with the
``MHD'' code for the equation of state, and E.\ Bica, P.\ Marigo, 
B.\ Salasnich and A.\ Weiss for the many useful discussions. 
L.\ Girardi acknowledges the
many people who, in the last years, either lended him computers and
CPU time in order to make preliminary calculations to the present
tracks (especially M.V.F.\ Copetti, A.A.\ Schmidt, S.O.\ Kepler), or
helped in the solution of daily computer problems (especially A.\
Weiss and B.\ Salasnich).  The work by L.\ Girardi has been initially 
funded by the CNPq Brazilian agency, and later by the German 
Alexander von Humboldt-Stiftung. This work was completed with 
funding by the Italian MURST.

\section*{ References }

\begin{description}
\item Alexander D.R., Ferguson J.W., 1994, ApJ 437, 879
\item Alongi M., Bertelli G., Bressan A., Chiosi C., 1991, A\&A
    244, 95 
\item Anders E., Grevesse N., 1989, Geochim.\ Cosmochim.\ Acta 53, 197
\item Aparicio A., Bertelli G., Chiosi C., Garcia-Pelayo J.M., 1990,
	A\&A 240, 262 
\item Baker N., Kippenhahn R., 1962, Z. Astroph. 54, 114
\item Bertelli G., Betto R., Bressan A., Chiosi C., Nasi E., 
      Vallenari A., 1990, A\&AS 85, 845
\item Bertelli G., Bressan A., Chiosi C., Angerer K., 1986, A\&AS 
	66, 191 
\item Bertelli G., Bressan A., Chiosi C., Fagotto F, Nasi E., 
      1994, A\&AS 106, 275
\item Bessell M.S. 1990, PASP 102, 1181
\item Bessell M.S., Brett J.M. 1988, PASP 100, 1134
\item B\"ohm-Vitense E., 1958, Z. Astroph. 46, 108
\item Bressan A., Bertelli G., Chiosi C., 1981, A\&A 102, 25 
\item Bressan A., Fagotto F., Bertelli G., Chiosi C., 1993, 
	A\&AS 100, 647
\item Buser R., Kurucz R.L., 1978, A\&A 70, 555
\item Carraro G., Vallenari A., Girardi L., Richichi A., 1999a, A\&A 
	343, 825
\item Carraro G., Girardi L., Chiosi C., 1999b, MNRAS 309, 430
\item Caughlan G.R., Fowler W.A., 1988, Atomic Data Nucl.\ Data Tables
	40, 283
\item Chiosi C., Bertelli G., Bressan A., 1992, ARA\&A 30, 305
\item Christensen-Dalsgaard J., Gough D.O., Thompson M.J., 1991, A\&A
	264, 518
\item Copeland H., Jensen J.O., Jorgensen H.E., 1970, A\&A 5, 12
\item D\"appen W., Mihalas D., Hummer D.G., Mihalas B.W., 1988, ApJ 332, 261
\item Fagotto F., Bressan A., Bertelli G., Chiosi C., 1994a, A\&AS 104, 365
\item Fagotto F., Bressan A., Bertelli G., Chiosi C., 1994b, A\&AS 105, 29
\item Girardi L., 1996, PhD thesis, Universidade Federal do Rio Grande
	do Sul, Porto Alegre, Brazil
\item Girardi L., Bressan A., Chiosi C., Bertelli G., Nasi E., 1996a,
	A\&AS 117, 113
\item Girardi L., Bressan A., Chiosi C., 1996b, in Stellar Evolution:
	What Should Be Done, 32nd Li\`ege Int.\ Astroph.\
	Coll., eds.\ A.\ Noels, D.\ Frapont-Caro, M.\ Gabriel, N.\ Grevesse,
	P.\ Demarque, p.\ 39
\item Girardi L., Bertelli G., 1998, MNRAS 300, 533
\item Girardi L., Groenewegen M.A.T., Weiss A., Salaris M., 1998,
	MNRAS 301, 149
\item Girardi L., 1999, MNRAS 308, 818
\item Graboske H.C., de Witt H.E., Grossman A.S., Cooper M.S., 1973,
        ApJ 181, 457
\item Grevesse N., Noels A., 1993, Phys. Scr. T, 47, 133
\item Groenewegen M.A.T., de Jong T., 1993, A\&A 267, 410
\item Herwig F., 1996, in Stellar Evolution: 
        What Should Be Done, 32nd Li\`ege Int.\ Astrophys.\ Coll.,
        eds.\ A.\ Noels et al., p.\ 441.
\item H\o g E., Pagel B.E.J., Portinari L., Thejll P.A., MacDonald J.,
	Girardi L., 1998, in Primordial Nuclei and their Galactic
	Evolution, eds.\ N.\ Prantzos, M.\ Tosi \& R.\ von Steiger,
	ISSI, Bern, Kluwer: Dordrecht, Space Science Reviews v.\ 84,
	p.\ 115
\item Hubbard W.B., Lampe M., 1969, ApJS 18, 297
\item Hummer D.G., Mihalas D., 1988, ApJ 331, 794 
\item Itoh N., Kohyama, 1983, ApJ 275, 858
\item Itoh N., Mitake S., Iyetomi H., Ichimaru S., 1983, ApJ 273, 774
\item Iglesias C.A., Rogers F.J., 1993, ApJ 412, 752
\item Jeffries R.D., 1997, MNRAS 288, 585
\item Kurucz R.L., 1992, in Stellar Populations of Galaxies, 
	eds.\ B.\ Barbuy and A.\ Renzini, Dordrecht: Kluwer, p.\ 225   
\item Landr\'e V., Prantzos N., Aguer P., Bogaert G., Lefebvre A., 
	Thibaud J.P., 1990, A\&A 240, 85
\item Maeder A., 1983, A\&A 120, 113
\item Marigo P., 1998a, PhD thesis, University of Padua, Italy
\item Marigo P., 1998b, A\&A 340, 463
\item Marigo P., Girardi L., Chiosi C., 1996, A\&A 316, L1
\item Marigo P., Bressan A., Chiosi C.\ 1998, A\&A 331, 564
\item Marigo P., Girardi L., Bressan A., 1999a, A\&A 344, 123
\item Marigo P., et al., 1999b, in preparation.
\item Meynet G., Maeder A., Schaller G., Schaerer D., Charbonnel C.,
	1994, A\&AS 103, 97
\item Mihalas D., D\"appen W., Hummer D.G., 1988, ApJ 331, 815
\item Mihalas D., Hummer D.G., Mihalas B.W., D\"appen W., 1990, ApJ
	350, 300
\item Munakata H., Kohyama Y., Itoh N., 1985, ApJ 296, 197
\item Pagel B.E.J., Portinari L., 1998, MNRAS 298, 747
\item Pasquini L., de Medeiros J.R., Girardi L., 1999, A\&A 
	in preparation. 
\item Perryman M.A.C., Lindegren L., Kovalevsky J., et al.,
	1995, A\&A 304, 69
\item Reimers D., 1975, Mem. Soc. R. Sci. Liege, ser. 6, vol. 8, p. 369 
\item Renzini A., Fusi Pecci F., 1988, ARA\&A 26, 199
\item Rogers F.J., Iglesias C.A., 1992, ApJS 79, 507
\item Salasnich B., Weiss A., Girardi L., Chiosi C., 1999, in preparation.
\item Straniero O., 1988, A\&AS 76, 157
\item Vassiliadis E., Wood P.R., 1993, ApJ 413, 641
\item Weaver T.A., Woosley S.E., 1993, Phys. Rep. 227, 65
\item Weidemann V., 1987, A\&A 188, 74
\item Weidemann V., 1996, in Advances in stellar evolution, eds.\ Rood
	R.T., Renzini A., p.\ 169
	\end{description}

\end{document}